\documentclass[prb,a4paper,twocolumn,floatfix,showpacs,showkeys,amsmath,amssymb,nobibnotes,unsortedaddress,superscriptaddress]{revtex4-2}
\setcitestyle{super}
\usepackage{graphicx}
\usepackage[export]{adjustbox}
\usepackage[dvipsnames]{xcolor}
\usepackage[utf8]{inputenc} 
\usepackage{amsmath}
\usepackage{xr}
\usepackage{upgreek}
\usepackage{subfigure}
\usepackage[percent]{overpic}
\usepackage{xcolor}

\graphicspath{{figures/}}

\begin{document}

\title{Power-law density of states in organic solar cells revealed by the open-circuit voltage dependence of the ideality factor}

\author{Maria Saladina}
\email[email: ]{maria.saladina@physik.tu-chemnitz.de}
\affiliation{Institut für Physik, Technische Universität Chemnitz, 09126 Chemnitz, Germany}

\author{Christopher Wöpke}
\affiliation{Institut für Physik, Technische Universität Chemnitz, 09126 Chemnitz, Germany}

\author{Clemens Göhler}
\affiliation{Institut für Physik, Technische Universität Chemnitz, 09126 Chemnitz, Germany}

\author{Ivan Ramirez}
\affiliation{Heliatek GmbH, 01139 Dresden, Germany}

\author{Olga Gerdes}
\affiliation{Heliatek GmbH, 01139 Dresden, Germany}

\author{Chao Liu}
\affiliation{Institute of Materials for Electronics and Energy Technology (i-MEET), Friedrich-Alexander-Universität Erlangen-Nürnberg, 91054 Erlangen, Germany}
\affiliation{Helmholtz Institute Erlangen-Nürnberg for Renewable Energy (HI ERN), 91058 Erlangen, Germany}

\author{Ning Li}
\affiliation{Institute of Materials for Electronics and Energy Technology (i-MEET), Friedrich-Alexander-Universität Erlangen-Nürnberg, 91054 Erlangen, Germany}
\affiliation{Helmholtz Institute Erlangen-Nürnberg for Renewable Energy (HI ERN), 91058 Erlangen, Germany}
\affiliation{State Key Laboratory of Luminescent Materials and Devices, Institute of Polymer Optoelectronic Materials and Devices, School of Materials Science and Engineering, South China University of Technology, 510640 Guangzhou, China}

\author{Thomas Heumüller}
\affiliation{Institute of Materials for Electronics and Energy Technology (i-MEET), Friedrich-Alexander-Universität Erlangen-Nürnberg, 91054 Erlangen, Germany}
\affiliation{Helmholtz Institute Erlangen-Nürnberg for Renewable Energy (HI ERN), 91058 Erlangen, Germany}

\author{Christoph J. Brabec}
\affiliation{Institute of Materials for Electronics and Energy Technology (i-MEET), Friedrich-Alexander-Universität Erlangen-Nürnberg, 91054 Erlangen, Germany}
\affiliation{Helmholtz Institute Erlangen-Nürnberg for Renewable Energy (HI ERN), 91058 Erlangen, Germany}

\author{Karsten Walzer}
\affiliation{Heliatek GmbH, 01139 Dresden, Germany}

\author{Martin Pfeiffer}
\affiliation{Heliatek GmbH, 01139 Dresden, Germany}

\author{Carsten Deibel}
\email[email: ]{deibel@physik.tu-chemnitz.de}
\affiliation{Institut für Physik, Technische Universität Chemnitz, 09126 Chemnitz, Germany}

\begin{abstract}

The density of states (DOS) is fundamentally important for understanding physical processes in organic disordered semiconductors, yet hard to determine experimentally. We evaluated the DOS by considering recombination via tail states and using the temperature and open-circuit voltage ($V_\mathrm{oc}$) dependence of the ideality factor in organic solar cells. By performing Suns-$V_\mathrm{oc}$ measurements, we find that gaussian and exponential distributions describe the DOS only at a given quasi-Fermi level splitting. The DOS width increases linearly with the DOS depth, revealing the power-law DOS in these materials.

\end{abstract}

\maketitle

\clearpage

The dominant recombination mechanism in a solar cell is intimately related to the ideality factor.\cite{wetzelaer2011trap,gohler2018nongeminate,tress2018interpretation} For inorganic semiconductors, the closer the ideality factor gets to 2, the more dominant the share of trap-assisted recombination.\cite{sah1957carrier} This connection is more complex for organic materials used in the state-of-the-art solar cells due to the energetic disorder inherent to these systems, giving rise to wave function localisation. Consequently, charge carrier transport and recombination in these disordered materials strongly dependent on the energetic distribution of localised states.\cite{bassler1993charge,rubel2004concentration,baranovskii2014theoretical,nenashev2015theoretical,hofacker2017dispersive} 
If the density of states (DOS) can be approximated by a gaussian, the ideality factor becomes unity and independent of temperature.\cite{hofacker2017dispersive,blakesley2011relationship} 
However, temperature-independent ideality factors equal to 1 are yet to be reported for organic donor--acceptor systems,\cite{street2012recombination,foertig2012shockley,tvingstedt2016temperature,perdigon2022understanding} 
implying that the DOS is more complicated in these materials.\cite{mackenzie2011modeling,mackenzie2012extracting,oelerich2012find} 

In this Letter, we seek to unravel the real shape of DOS in a set of solar cells based on organic semiconductor blends. To achieve this objective, we determine ideality factors from temperature-dependent Suns-$V_\mathrm{oc}$ measurements. We connect the results to theoretical predictions by the multiple-trapping-and-release (MTR) model\cite{baranovskii2014theoretical,nenashev2015theoretical,noolandi1977multiple,arkhipov1984dispersive} using different combinations of the gaussian and exponential DOS functions\cite{mark1962space,paasch2010charge,garcia2010temperature,kirchartz2012meaning,burke2015beyond,xiao2020relationship} to describe the energetic state distribution of electrons and holes. 
Depending on the shape of DOS and the dominant recombination mechanism, the temperature dependence of the ideality factor differs,\cite{hofacker2017dispersive} which allows us to assign specific recombination models to the investigated systems. We find that the width of the DOS distribution in these organic solar cells depends on the energetic position in the DOS, resulting in a power-law distribution of localised states.

Current--voltage characteristics of a solar cell are usually approximated by the diode equation.\cite{shockley1949theory} The recombination rate $R$ enters the diode equation via recombination current density $j_\mathrm{rec}$, which at the open-circuit conditions takes the form \cite{wurfel2015impact}
\begin{equation}\label{eq:01}\begin{split}
    j_\mathrm{rec} &= e \int_{0}^{L} R(x) \,dx \approx eLR \\
    &= j_0\cdot\exp{\left(\frac{eV_\mathrm{oc}}{n_\mathrm{id} k_BT}\right)} . 
\end{split}\end{equation}
Here $L$ stands for the active layer thickness, $j_0$ the dark saturation current density, $n_\mathrm{id}$ the ideality factor, $V_\mathrm{oc}$ the open-circuit voltage, $e$ the elementary charge, $k_B$ the Boltzmann constant, and $T$ the temperature. In the framework of the extended Onsager model,\cite{noolandi1979theory,nikitenko2001dispersive,hilczer2010unified} recombination takes place via the formation of a charge-transfer (CT) state, when charge carriers are within a certain distance of one another. Consequently, for recombination to occur at the donor--acceptor interface, at least one of the charge carriers needs to be mobile, and recombination of trapped charge carriers with each other is excluded.\cite{tachiya2010theory,gorenflot2014nongeminate} Then, the recombination rate is expressed as
\begin{equation}\label{eq:02}
    R = k_{r,np} n_c p_c + k_{r,n} n_c p_t + k_{r,p} n_t p_c , 
\end{equation}
where $k_r$ stands for the recombination prefactor, and the subscript denotes which of the charge carriers, electrons $n$ or holes $p$, are mobile. We use different prefactors to highlight the dependence of $k_r$ on the combined mobility of charge carriers. The finite escape probability from CT back to the separated state, if present, is included in $k_r$ and reduces it compared to the Langevin prefactor $k_L = e \left( \mu_n + \mu_p \right)/\varepsilon$.\cite{langevin1903recombinaison,braun1984electric,shoaee2019decoding} Additionally, $k_r$ contains the effects of active layer morphology that cause its further deviation from $k_L$.\cite{koster2006bimolecular,heiber2015encounter,heiber2016charge}

In organic disordered materials, the majority of localised states in the DOS lie below the transport energy,\cite{arkhipov2001effective} and act as traps, which capture mobile charges. Trapped charge carriers can be thermally released and contribute to photoconductivity. During this process of multiple-trapping-and-release,\cite{baranovskii2014theoretical,nenashev2015theoretical,noolandi1977multiple,arkhipov1984dispersive} some share of charge carriers recombines and is lost to the photocurrent. 
In the MTR model, the fraction of the mobile charge carrier density is expressed through parameter $\theta$, the trapping factor, which depends on the DOS distribution.\cite{baranovskii2014theoretical,arkhipov1984dispersive,adriaenssens1995photoconductivity} 
The density of mobile and trapped charge carriers is expressed as $n_c = \theta n$ and $n_t = \left( 1 - \theta \right) n$, respectively, with $\theta < 1$. As the effective density of trap states is much larger than the charge carrier density, most relaxed charge carriers will populate energy sites in the DOS tail.\cite{arkhipov2005charge} Thus, $n_c \ll n_t \approx n$ and $\theta \ll 1$, inferring that recombination is mainly trap-mediated and making the first term in Eq.\,\eqref{eq:02} negligible. Using the above notations, and noting that due to the nature of photogeneration, $n=p$, the recombination rate becomes 
\begin{equation}\label{eq:03}
    R \approx \left( k_{r,n} \theta_n + k_{r,p} \theta_p \right) n^2 . 
\end{equation}

The two recombination channels in Eq.\,\eqref{eq:03} are distinguished by the type of mobile charge carrier. One of the channels is dominant if its recombination prefactor and/or its trapping factor is larger than for the other channel. Thus, the exact expression of $R$ depends on (i) the physical parameters, e.g.\ mobility, of the mobile charge carrier type through the recombination prefactor $k_r$, (ii) the DOS of this charge carrier type through the trapping factor $\theta$, and (iii) the DOS of the more abundant type of charge carrier in the dominant recombination channel through the total charge carrier concentration $n$.

Herein, we focus on the most prevalent models used to approximate the DOS distributions in organic semiconductors -- the gaussian and exponential DOS,\cite{mark1962space,paasch2010charge,garcia2010temperature,kirchartz2012meaning,burke2015beyond,xiao2020relationship} and their influence on the ideality factor. The depth of trap states, corresponding to the width of the distribution, depends on the disorder parameter $\sigma$ and the Urbach energy $E_\mathrm{U}$, respectively. 
The resulting form of Eq.\,\eqref{eq:03} is defined by four combinations of these DOS distributions. The first two involve electrons and holes being described by the same DOS distribution, whether gaussian, or exponential, and, for the sake of simplicity, $\theta_n=\theta_p$. If, however, the DOS functions of electrons and holes are different, and $\theta_n\neq\theta_p$, the effective recombination rate is additionally determined by the type of mobile charge carrier. 
For the detailed derivation, the interested reader is referred to the comprehensive work of Hofacker and Neher. \cite{hofacker2017dispersive} Here, we build on a mere fraction of their results related to the ideality factor and summarise relevant parts of the derivation in the Supplemental Material. The ideality factor is obtained by comparing Eq.\,\eqref{eq:01} to the equations of $R$ for the DOS combinations discussed above (Eqs.\,(S11) to (S14)).

Without loss of generality, we describe the dominant recombination channel involving mobile holes recombining with trapped electrons. 
The more abundant type of charge carrier in the recombination channel ($n_t \approx n$) controls the temperature dependence of the ideality factor, while the mobile charge carrier type ($p_c = \theta_p \cdot p$) controls the recombination order. 
If the DOS of electrons is described by an exponential, the ideality factor is temperature-dependent. When such electrons recombine with mobile holes from the gaussian DOS, the ideality factor is independent of $\sigma$ and is expressed as\cite{hofacker2017dispersive}
\begin{equation}\label{eq:04}
    n_\mathrm{id} = \frac{E_\mathrm{U} + k_BT}{2k_BT} , 
\end{equation} 
If mobile holes are also represented by the exponential DOS, the ideality factor is given by\cite{hofacker2017dispersive,foertig2012shockley,van1993quality,kirchartz2011recombination}
\begin{equation}\label{eq:05}
    n_\mathrm{id} = \frac{2E_\mathrm{U}}{E_\mathrm{U} + k_BT} . 
\end{equation}
We will be referring to these models as the \textit{mixed DOS} and \textit{exponential DOS}, respectively. 
In contrast, if the DOS of electrons is described by a gaussian, in the low concentration limit we arrive at $n_\mathrm{id} = 1$, independent of temperature.\cite{hofacker2017dispersive,blakesley2011relationship,pasveer2005unified} This is true irrespective of whether mobile holes come from the gaussian or exponential DOS.

An ideality factor of unity is not observed experimentally in organic semiconductors,\cite{street2012recombination,foertig2012shockley,tvingstedt2016temperature,perdigon2022understanding} which leads to two implications. Firstly, in a mixed DOS, the dominant recombination channel is the one involving mobile charge carriers in the gaussian recombining with trapped charges in the exponential DOS. 
A gaussian DOS reaches less deep into the band gap so that $\theta$ is generally closer to one than for an exponential DOS. Hence, this channel will have a larger share of mobile charge carriers leading, for the same $k_r$, to a larger effective recombination prefactor than for the other channel. Secondly, the total distribution of localised states is likely more complicated than the gaussian for organic materials. 

\begin{figure}[t]\centering
    \includegraphics[width=0.4\textwidth]{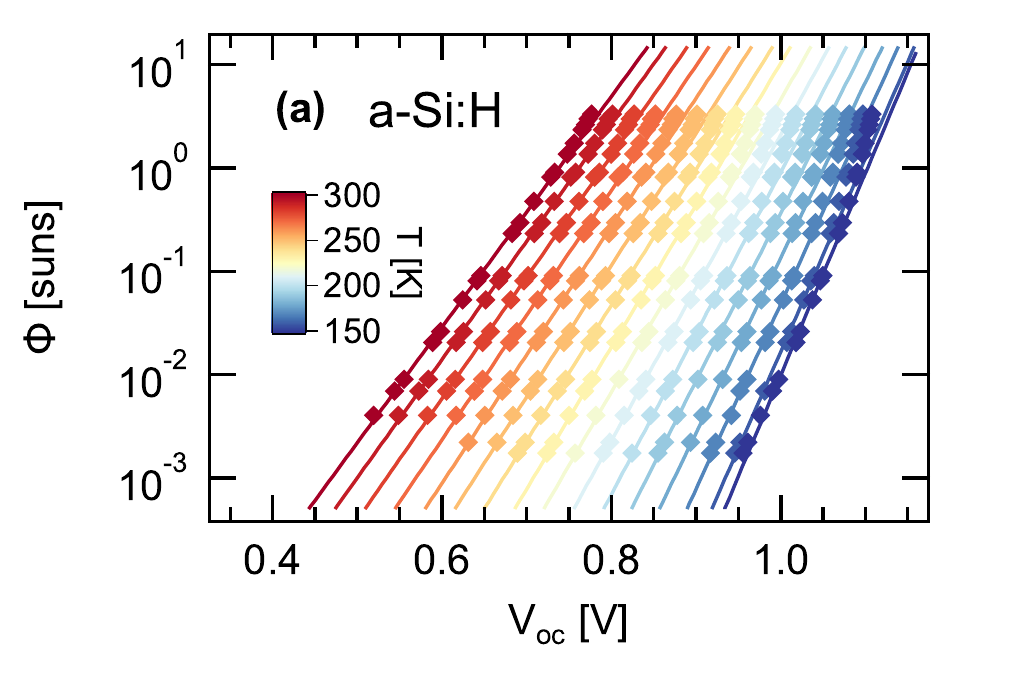}
    \includegraphics[width=0.4\textwidth]{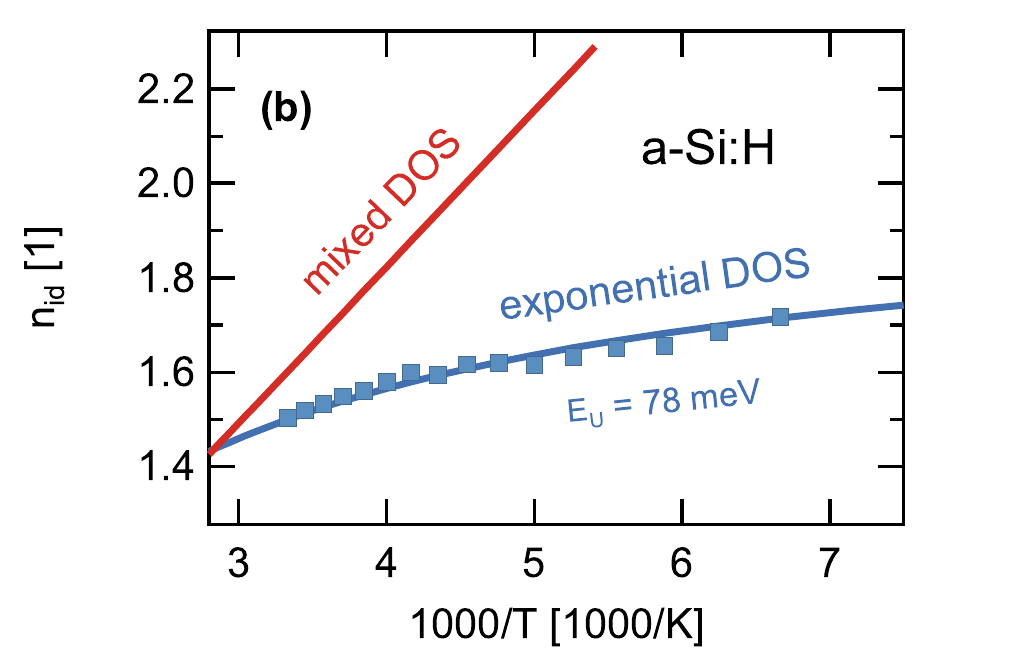}
    \caption{(a) Suns-$V_\mathrm{oc}$ data (symbols) of a-Si:H solar cell fitted with Eq.\,\eqref{eq:06} (solid lines). (b) Temperature-dependent ideality factor $n_\mathrm{id}$ extracted from the fits in comparison to Eqs.\,\eqref{eq:04} and \eqref{eq:05}.}
\label{fig:01}\end{figure}

Consequently, in order to shed light on the shape of the DOS in these systems, our focus should lie on the temperature dependence of the ideality factor, with the models underlying Eqs.\,\eqref{eq:04} and \eqref{eq:05} as the starting point. The distinct temperature dependence of $n_\mathrm{id}$ in these expressions allows us to determine the prevailing recombination mechanism in a solar cell dominated by trap-assisted recombination, and the likely form of DOS distribution.

To verify that the DOS can be established through ideality factors, we chose the well-studied hydrogenated amorphous silicon solar cell as a reference. We then expand our investigation to a set of material systems representative of typical organic solar cell classes, such as solution-processed fullerene (P3HT:PC$_{61}$BM) and non-fullerene acceptor devices (PM6:Y6), along with thermally evaporated small-molecule solar cells (DCV-V-Fu-Ind-Fu-V:C$_{60}$). 
The details of molecular structure and device fabrication are given in the Supplemental Material. 

We employ illumination intensity-dependent $V_\mathrm{oc}$ measurements to determine ideality factors in the absence of series and transport resistance.\cite{wolf1963series,kirchartz2013differences,wopke2022traps} Figure~\ref{fig:01}(a) shows the Suns-$V_\mathrm{oc}$ data of a-Si:H solar cell between 150\,K and 300\,K, excluding the regions of low light intensity influenced by low shunt resistance. Roughly above 1\,sun, $V_\mathrm{oc}$ becomes limited by the contacts, which is more pronounced at low temperatures. 
Ideality factors were extracted from the slope of $\Phi(V_\mathrm{oc})$ according to \cite{tvingstedt2016temperature}
\begin{equation}\label{eq:06}
    n_\mathrm{id} = \frac{e}{k_BT}\left(\frac{d\ln{\Phi}}{dV_\mathrm{oc}}\right)^{-1} . 
\end{equation}
For each temperature, the data can be fitted with a single slope over two orders of magnitude of light intensities. 

We plot the resulting ideality factors against the inverse temperature in Figure~\ref{fig:01}(b). Consistent with van Berkel et al.,\cite{van1993quality} we observe the decrease of $n_\mathrm{id}$ of a-Si:H solar cell with higher temperature from 1.7 at 150\,K to 1.5 at 300\,K. 
The temperature dependence of $n_\mathrm{id}$ can be fitted with Eq.\,\eqref{eq:05} and therefore is assigned to the trap-assisted recombination of charge carriers with the exponential density of states. In the exponential DOS model $n_\mathrm{id} \propto \left( E_\mathrm{U} + k_BT \right)^{-1}$, resulting in sublinear dependence on $1/T$. The fit yields the Urbach energy of $\approx 78$\,meV, in agreement with the literature,\cite{van1993quality,wehrspohn2000relative} which is independent of temperature and light intensity. 
A mixed DOS would lead to a distinctly different temperature dependence of the ideality factor, as $n_\mathrm{id} \propto 1/T$ according to Eq.\,\eqref{eq:04}. 

\begin{figure}[b]\centering
    \includegraphics[width=0.4\textwidth]{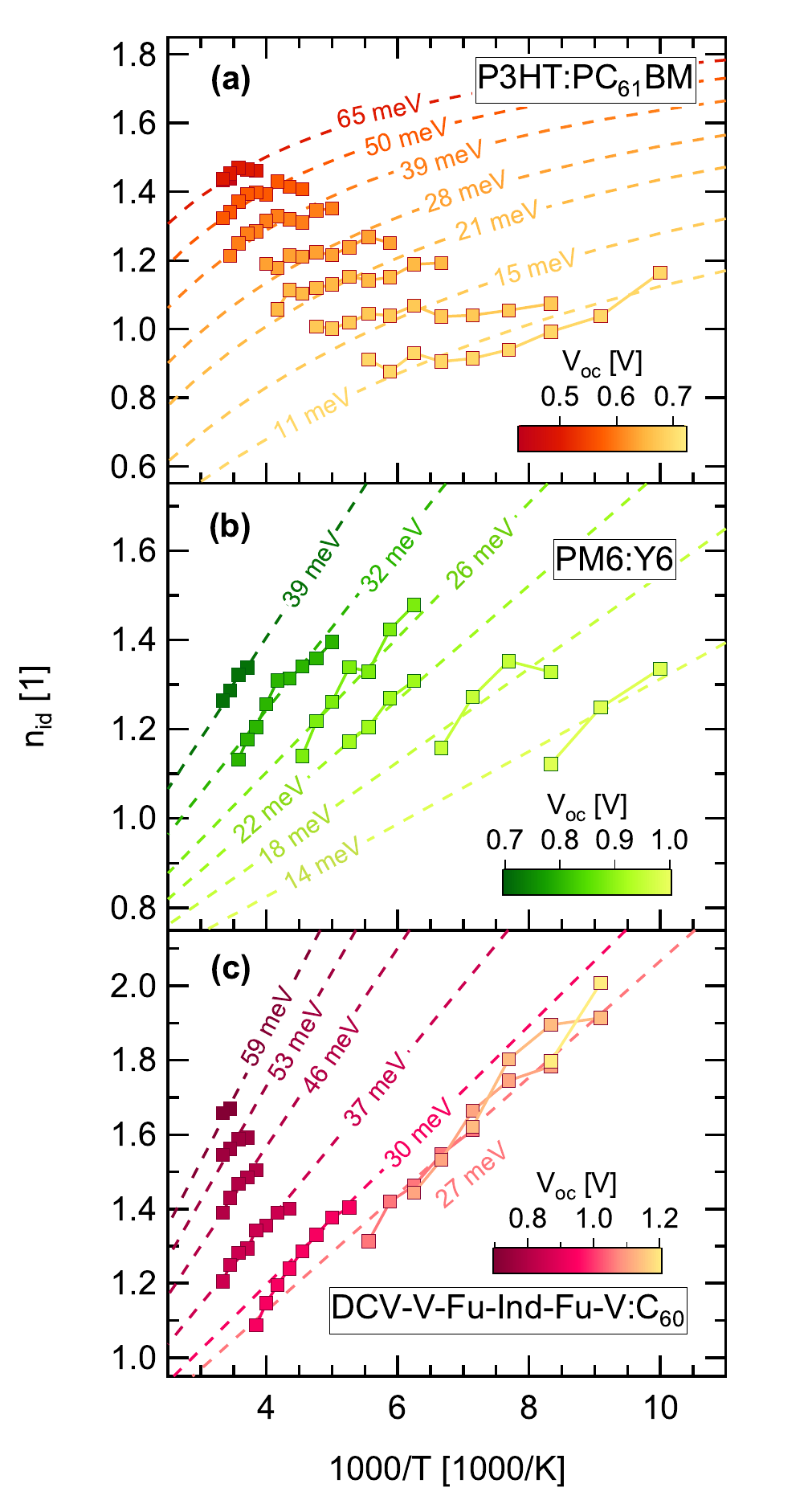}
    \caption{Temperature-dependent ideality factors $n_\mathrm{id}$ of (a) P3HT:PC$_{61}$BM, (b) PM6:Y6 and (c) DCV-V-Fu-Ind-Fu-V:C$_{60}$ (symbols). Darker color corresponds to lower $V_\mathrm{oc}$, i.e.\ deeper subgap energy states. Dashed lines are the calculated $n_\mathrm{id}(E_\mathrm{U},T)$ according to Eq.\,\eqref{eq:05} for P3HT:PC$_{61}$BM and Eq.\,\eqref{eq:04} for PM6:Y6 and DCV-V-Fu-Ind-Fu-V:C$_{60}$. $E_\mathrm{U}$ increases with the DOS depth for all three systems.}
\label{fig:02}\end{figure}

We now extend the scope of the study to organic donor--acceptor systems. 
Figure~\ref{fig:02} shows ideality factors of P3HT:PC$_{61}$BM, PM6:Y6 and DCV-V-Fu-Ind-Fu-V:C$_{60}$.  
First, we note that at each temperature $n_\mathrm{id}$ has several values corresponding to the local slope of $\Phi(V_\mathrm{oc})$. 
Hence, in contrast to a-Si:H, the ideality factor of the organic systems we investigate here is light intensity-dependent, and generally decreases with increasing light intensity. 
At first, it seems problematic to assign a specific recombination model based on the temperature dependence of $n_\mathrm{id}$ at a certain illumination intensity (cf.\ Figure~S3). This method does not account for potential changes of DOS shape with the quasi-Fermi level splitting (QFLS), which varies with temperature for fixed light intensity. 

\begin{figure*}[!ht]\centering
    \includegraphics[width=0.8\textwidth]{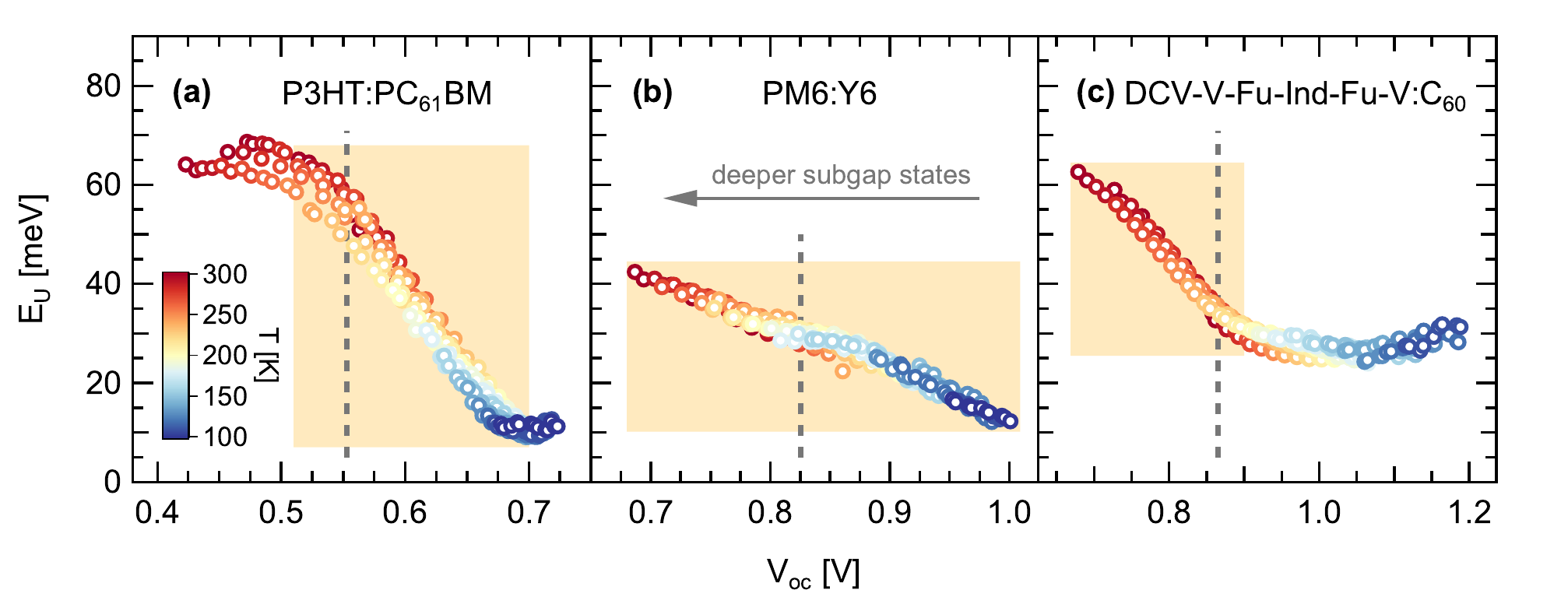}
    \caption{Urbach energies $E_\mathrm{U}$ as a function of the quasi-Fermi level splitting, approximated by $V_\mathrm{oc}$. $E_\mathrm{U}$ was calculated from $n_\mathrm{id}(T)$ according to Eq.\,\eqref{eq:05} for P3HT:PC$_{61}$BM and Eq.\,\eqref{eq:04} for PM6:Y6 and DCV-V-Fu-Ind-Fu-V:C$_{60}$. $E_\mathrm{U}$ is independent of $T$, while it depends linearly on $V_\mathrm{oc}$ within the highlighted areas, inferring that the DOS is described by a power law. Roughly constant $E_\mathrm{U}$ outside of the highlighted areas is an indication of exponential DOS. Dashed lines correspond to $V_\mathrm{oc}$ at 1\,sun illumination intensity at room temperature.}
\label{fig:03}\end{figure*}

Instead, we evaluate the data at fixed $V_\mathrm{oc}$. 
The QFLS, approximated by $V_\mathrm{oc}$, samples the combined DOS of electrons and holes at a certain energy. The Urbach energy, which is a measure of disorder for the exponential DOS distribution, is independent of temperature at fixed $V_\mathrm{oc}$, and the ideality factor describes the dominant recombination mechanism at this DOS depth. Coming back to Figure\,\ref{fig:02}, we see our approach paying off. For a given $V_\mathrm{oc}$ the relation between $n_\mathrm{id}$ and $1/T$ can be assigned to specific recombination models for all three material systems. 

Ideality factors of P3HT:PC$_{61}$BM in Figure\,\ref{fig:02}(a) at a given $V_\mathrm{oc}$ follow the exponential DOS model, Eq.\,\eqref{eq:05}, similar to what we found for a-Si:H earlier. PM6:Y6 and DCV-V-Fu-Ind-Fu-V:C$_{60}$ in Figures\,\ref{fig:02}(b) and (c), on the other hand, are best described by the mixed DOS model, Eq.\,\eqref{eq:04}. 
For all three organic donor--acceptor systems, the Urbach energy depends on the QFLS. 
We calculate ideality factors using different values of $E_\mathrm{U}$ (indicated by dashed lines in the figure) and find good agreement with the data, inferring that the exponential DOS acts as a local approximation of the real DOS at a given QFLS.\cite{mackenzie2012extracting} 
The deeper we are probing in the density of states, the larger the $E_\mathrm{U}$. The Urbach energy of the thermally evaporated DCV-V-Fu-Ind-Fu-V:C$_{60}$ solar cell is roughly constant above 0.90\,V, as $n_\mathrm{id}$ converges at different $V_\mathrm{oc}$, suggesting that in this region the DOS is nearly exponential. 

Figure\,\ref{fig:03} shows the relation between the Urbach energy and the QFLS. First, we note a striking convergence of $E_\mathrm{U}$ at different temperatures for all the systems, solidifying our confidence in the analytical approach and the assigned system-specific recombination models. 
$E_\mathrm{U}$ depends linearly on the QFLS in the regions of non-constant Urbach energy (highlighted areas in the figure). This linearity helps to unravel the real shape of the density of states. For the exponential DOS $g_{\text{exp}}(E)$
\begin{equation}\label{eq:07}\begin{split}
    \frac{\mathrm{d}\ln{g_{\text{exp}}(E)}}{\mathrm{d}E} &= \frac{1}{E_\mathrm{U}} . 
\end{split}\end{equation}
We emphasise again that in our case exponential function acts as a local approximation of the real DOS. With the linear relation that we observe, $E_\mathrm{U}(E) \propto E/\xi$, the derivative of the real DOS distribution is given by 
\begin{equation}\label{eq:08}\begin{split}
    \frac{\mathrm{d}\ln{g_{\text{}}(E)}}{\mathrm{d}E} &\propto \frac{\xi}{E} . 
\end{split}\end{equation}
This relation leads to $\ln{g_{\text{}}(E)} \propto \xi\ln{E}$, and the ultimate form of the DOS distribution is a power-law
\begin{equation}\label{eq:09}\begin{split}
    g_{\text{}}(E) &\propto E^\xi . \\
\end{split}\end{equation}

To our knowledge, the power-law density of states has not been reported for organic solar cells before. However, it must be considered when explaining experimental data and simulating the physics of organic semiconductors, as we observe this DOS distribution in all the donor--acceptor systems investigated herein. The power-law DOS can be assigned to P3HT:PC$_{61}$BM in the $V_\mathrm{oc}$ range between ca.\ 0.51\,V and 0.70\,V, and to PM6:Y6 in the whole measured data range. The density of states of DCV-V-Fu-Ind-Fu-V:C$_{60}$ follows the power law up to ca.\ 0.90\,V and changes its shape to exponential at higher $V_\mathrm{oc}$. Compared to the exponential DOS, the power law is described by a narrower tail. Hence, there are fewer available states for the charge carriers to fill up, resulting in a higher $V_\mathrm{oc}$ than would be expected for the exponential DOS. 

By averaging, we find that under 1 sun illumination at room temperature $E_\mathrm{U} \approx 55$\,meV for P3HT:PC$_{61}$BM, in good agreement with previous reports.\cite{foertig2012shockley,kirchartz2011recombination} 
In DCV-V-Fu-Ind-Fu-V:C$_{60}$ the Urbach energy $\approx 35$\,meV at 0.87\,V, corresponding to 1 sun illumination. This implies that the trap distribution is narrower in this system compared to P3HT:PC$_{61}$BM. 
In PM6:Y6 energetic disorder is reduced even further, as $E_\mathrm{U}$ in this system is only 30\,meV, consistent with the literature values.\cite{karki2019understanding,wu2020exceptionally,yang2020compatible}
Lower disorder is in general beneficial to the solar cell performance, as it leads to enhanced charge carrier transport and reduced $V_\mathrm{oc}$ losses.\cite{lange2013correlation,collins2016understanding,goehler2022role} 

In conclusion, we employ temperature and illumination intensity-dependent $V_\mathrm{oc}$ measurements to determine the type of the density of states in PM6:Y6, DCV-V-Fu-Ind-Fu-V:C$_{60}$ and P3HT:PC$_{61}$BM organic solar cells. 
We find that the temperature dependence of the ideality factor can be explained in the framework of the multiple-trapping-and-release model, but only if analysed at a given open-circuit voltage. 
For all the investigated systems, the density of states of trapped charges participating in recombination is locally approximated by the exponential one, where the width of the distribution decreases linearly with increasing quasi-Fermi level splitting. Our results establish that the density of states in these solar cells follows a power law, which has not been reported for organic donor--acceptor systems up to date. We suggest that the temperature-dependent experiments designed to understand recombination mechanisms in these disordered systems have to be conducted at the same quasi-Fermi level splitting to ensure correct interpretation of the results. 

The authors are grateful to J.\ Gorenflot, KAUST, for his valuable comments on the manuscript. The work was partly funded by the DFG (projects DE 830/19-1 and DE 830/26-1).

\bibliographystyle{apsrev4-2}
\bibliography{main}

\begin{thebibliography}{54}%
\makeatletter
\providecommand \@ifxundefined [1]{%
 \@ifx{#1\undefined}
}%
\providecommand \@ifnum [1]{%
 \ifnum #1\expandafter \@firstoftwo
 \else \expandafter \@secondoftwo
 \fi
}%
\providecommand \@ifx [1]{%
 \ifx #1\expandafter \@firstoftwo
 \else \expandafter \@secondoftwo
 \fi
}%
\providecommand \natexlab [1]{#1}%
\providecommand \enquote  [1]{``#1''}%
\providecommand \bibnamefont  [1]{#1}%
\providecommand \bibfnamefont [1]{#1}%
\providecommand \citenamefont [1]{#1}%
\providecommand \href@noop [0]{\@secondoftwo}%
\providecommand \href [0]{\begingroup \@sanitize@url \@href}%
\providecommand \@href[1]{\@@startlink{#1}\@@href}%
\providecommand \@@href[1]{\endgroup#1\@@endlink}%
\providecommand \@sanitize@url [0]{\catcode `\\12\catcode `\$12\catcode
  `\&12\catcode `\#12\catcode `\^12\catcode `\_12\catcode `\%12\relax}%
\providecommand \@@startlink[1]{}%
\providecommand \@@endlink[0]{}%
\providecommand \url  [0]{\begingroup\@sanitize@url \@url }%
\providecommand \@url [1]{\endgroup\@href {#1}{\urlprefix }}%
\providecommand \urlprefix  [0]{URL }%
\providecommand \Eprint [0]{\href }%
\providecommand \doibase [0]{https://doi.org/}%
\providecommand \selectlanguage [0]{\@gobble}%
\providecommand \bibinfo  [0]{\@secondoftwo}%
\providecommand \bibfield  [0]{\@secondoftwo}%
\providecommand \translation [1]{[#1]}%
\providecommand \BibitemOpen [0]{}%
\providecommand \bibitemStop [0]{}%
\providecommand \bibitemNoStop [0]{.\EOS\space}%
\providecommand \EOS [0]{\spacefactor3000\relax}%
\providecommand \BibitemShut  [1]{\csname bibitem#1\endcsname}%
\let\auto@bib@innerbib\@empty
\bibitem [{\citenamefont {Wetzelaer}\ \emph {et~al.}(2011)\citenamefont
  {Wetzelaer}, \citenamefont {Kuik}, \citenamefont {Nicolai},\ and\
  \citenamefont {Blom}}]{wetzelaer2011trap}%
  \BibitemOpen
  \bibfield  {author} {\bibinfo {author} {\bibfnamefont {G.}~\bibnamefont
  {Wetzelaer}}, \bibinfo {author} {\bibfnamefont {M.}~\bibnamefont {Kuik}},
  \bibinfo {author} {\bibfnamefont {H.}~\bibnamefont {Nicolai}},\ and\ \bibinfo
  {author} {\bibfnamefont {P.}~\bibnamefont {Blom}},\ }\href@noop {} {\bibfield
   {journal} {\bibinfo  {journal} {Physical Review B}\ }\textbf {\bibinfo
  {volume} {83}},\ \bibinfo {pages} {165204} (\bibinfo {year}
  {2011})}\BibitemShut {NoStop}%
\bibitem [{\citenamefont {G{\"o}hler}\ \emph {et~al.}(2018)\citenamefont
  {G{\"o}hler}, \citenamefont {Wagenpfahl},\ and\ \citenamefont
  {Deibel}}]{gohler2018nongeminate}%
  \BibitemOpen
  \bibfield  {author} {\bibinfo {author} {\bibfnamefont {C.}~\bibnamefont
  {G{\"o}hler}}, \bibinfo {author} {\bibfnamefont {A.}~\bibnamefont
  {Wagenpfahl}},\ and\ \bibinfo {author} {\bibfnamefont {C.}~\bibnamefont
  {Deibel}},\ }\href@noop {} {\bibfield  {journal} {\bibinfo  {journal}
  {Advanced Electronic Materials}\ }\textbf {\bibinfo {volume} {4}},\ \bibinfo
  {pages} {1700505} (\bibinfo {year} {2018})}\BibitemShut {NoStop}%
\bibitem [{\citenamefont {Tress}\ \emph {et~al.}(2018)\citenamefont {Tress},
  \citenamefont {Yavari}, \citenamefont {Domanski}, \citenamefont {Yadav},
  \citenamefont {Niesen}, \citenamefont {Baena}, \citenamefont {Hagfeldt},\
  and\ \citenamefont {Graetzel}}]{tress2018interpretation}%
  \BibitemOpen
  \bibfield  {author} {\bibinfo {author} {\bibfnamefont {W.}~\bibnamefont
  {Tress}}, \bibinfo {author} {\bibfnamefont {M.}~\bibnamefont {Yavari}},
  \bibinfo {author} {\bibfnamefont {K.}~\bibnamefont {Domanski}}, \bibinfo
  {author} {\bibfnamefont {P.}~\bibnamefont {Yadav}}, \bibinfo {author}
  {\bibfnamefont {B.}~\bibnamefont {Niesen}}, \bibinfo {author} {\bibfnamefont
  {J.~P.~C.}\ \bibnamefont {Baena}}, \bibinfo {author} {\bibfnamefont
  {A.}~\bibnamefont {Hagfeldt}},\ and\ \bibinfo {author} {\bibfnamefont
  {M.}~\bibnamefont {Graetzel}},\ }\href@noop {} {\bibfield  {journal}
  {\bibinfo  {journal} {Energy \& Environmental Science}\ }\textbf {\bibinfo
  {volume} {11}},\ \bibinfo {pages} {151} (\bibinfo {year} {2018})}\BibitemShut
  {NoStop}%
\bibitem [{\citenamefont {Sah}\ \emph {et~al.}(1957)\citenamefont {Sah},
  \citenamefont {Noyce},\ and\ \citenamefont {Shockley}}]{sah1957carrier}%
  \BibitemOpen
  \bibfield  {author} {\bibinfo {author} {\bibfnamefont {C.-T.}\ \bibnamefont
  {Sah}}, \bibinfo {author} {\bibfnamefont {R.~N.}\ \bibnamefont {Noyce}},\
  and\ \bibinfo {author} {\bibfnamefont {W.}~\bibnamefont {Shockley}},\
  }\href@noop {} {\bibfield  {journal} {\bibinfo  {journal} {Proceedings of the
  IRE}\ }\textbf {\bibinfo {volume} {45}},\ \bibinfo {pages} {1228} (\bibinfo
  {year} {1957})}\BibitemShut {NoStop}%
\bibitem [{\citenamefont {B{\"a}ssler}(1993)}]{bassler1993charge}%
  \BibitemOpen
  \bibfield  {author} {\bibinfo {author} {\bibfnamefont {H.}~\bibnamefont
  {B{\"a}ssler}},\ }\href@noop {} {\bibfield  {journal} {\bibinfo  {journal}
  {Physica Status Solidi (b)}\ }\textbf {\bibinfo {volume} {175}},\ \bibinfo
  {pages} {15} (\bibinfo {year} {1993})}\BibitemShut {NoStop}%
\bibitem [{\citenamefont {Rubel}\ \emph {et~al.}(2004)\citenamefont {Rubel},
  \citenamefont {Baranovskii}, \citenamefont {Thomas},\ and\ \citenamefont
  {Yamasaki}}]{rubel2004concentration}%
  \BibitemOpen
  \bibfield  {author} {\bibinfo {author} {\bibfnamefont {O.}~\bibnamefont
  {Rubel}}, \bibinfo {author} {\bibfnamefont {S.}~\bibnamefont {Baranovskii}},
  \bibinfo {author} {\bibfnamefont {P.}~\bibnamefont {Thomas}},\ and\ \bibinfo
  {author} {\bibfnamefont {S.}~\bibnamefont {Yamasaki}},\ }\href@noop {}
  {\bibfield  {journal} {\bibinfo  {journal} {Physical Review B}\ }\textbf
  {\bibinfo {volume} {69}},\ \bibinfo {pages} {014206} (\bibinfo {year}
  {2004})}\BibitemShut {NoStop}%
\bibitem [{\citenamefont {Baranovskii}(2014)}]{baranovskii2014theoretical}%
  \BibitemOpen
  \bibfield  {author} {\bibinfo {author} {\bibfnamefont {S.}~\bibnamefont
  {Baranovskii}},\ }\href@noop {} {\bibfield  {journal} {\bibinfo  {journal}
  {Physica Status Solidi (b)}\ }\textbf {\bibinfo {volume} {251}},\ \bibinfo
  {pages} {487} (\bibinfo {year} {2014})}\BibitemShut {NoStop}%
\bibitem [{\citenamefont {Nenashev}\ \emph {et~al.}(2015)\citenamefont
  {Nenashev}, \citenamefont {Oelerich},\ and\ \citenamefont
  {Baranovskii}}]{nenashev2015theoretical}%
  \BibitemOpen
  \bibfield  {author} {\bibinfo {author} {\bibfnamefont {A.}~\bibnamefont
  {Nenashev}}, \bibinfo {author} {\bibfnamefont {J.}~\bibnamefont {Oelerich}},\
  and\ \bibinfo {author} {\bibfnamefont {S.}~\bibnamefont {Baranovskii}},\
  }\href@noop {} {\bibfield  {journal} {\bibinfo  {journal} {Journal of
  Physics: Condensed Matter}\ }\textbf {\bibinfo {volume} {27}},\ \bibinfo
  {pages} {093201} (\bibinfo {year} {2015})}\BibitemShut {NoStop}%
\bibitem [{\citenamefont {Hofacker}\ and\ \citenamefont
  {Neher}(2017)}]{hofacker2017dispersive}%
  \BibitemOpen
  \bibfield  {author} {\bibinfo {author} {\bibfnamefont {A.}~\bibnamefont
  {Hofacker}}\ and\ \bibinfo {author} {\bibfnamefont {D.}~\bibnamefont
  {Neher}},\ }\href@noop {} {\bibfield  {journal} {\bibinfo  {journal}
  {Physical Review B}\ }\textbf {\bibinfo {volume} {96}},\ \bibinfo {pages}
  {245204} (\bibinfo {year} {2017})}\BibitemShut {NoStop}%
\bibitem [{\citenamefont {Blakesley}\ and\ \citenamefont
  {Neher}(2011)}]{blakesley2011relationship}%
  \BibitemOpen
  \bibfield  {author} {\bibinfo {author} {\bibfnamefont {J.~C.}\ \bibnamefont
  {Blakesley}}\ and\ \bibinfo {author} {\bibfnamefont {D.}~\bibnamefont
  {Neher}},\ }\href@noop {} {\bibfield  {journal} {\bibinfo  {journal}
  {Physical Review B}\ }\textbf {\bibinfo {volume} {84}},\ \bibinfo {pages}
  {075210} (\bibinfo {year} {2011})}\BibitemShut {NoStop}%
\bibitem [{\citenamefont {Street}\ \emph {et~al.}(2012)\citenamefont {Street},
  \citenamefont {Krakaris},\ and\ \citenamefont
  {Cowan}}]{street2012recombination}%
  \BibitemOpen
  \bibfield  {author} {\bibinfo {author} {\bibfnamefont {R.~A.}\ \bibnamefont
  {Street}}, \bibinfo {author} {\bibfnamefont {A.}~\bibnamefont {Krakaris}},\
  and\ \bibinfo {author} {\bibfnamefont {S.~R.}\ \bibnamefont {Cowan}},\
  }\href@noop {} {\bibfield  {journal} {\bibinfo  {journal} {Advanced
  Functional Materials}\ }\textbf {\bibinfo {volume} {22}},\ \bibinfo {pages}
  {4608} (\bibinfo {year} {2012})}\BibitemShut {NoStop}%
\bibitem [{\citenamefont {Foertig}\ \emph {et~al.}(2012)\citenamefont
  {Foertig}, \citenamefont {Rauh}, \citenamefont {Dyakonov},\ and\
  \citenamefont {Deibel}}]{foertig2012shockley}%
  \BibitemOpen
  \bibfield  {author} {\bibinfo {author} {\bibfnamefont {A.}~\bibnamefont
  {Foertig}}, \bibinfo {author} {\bibfnamefont {J.}~\bibnamefont {Rauh}},
  \bibinfo {author} {\bibfnamefont {V.}~\bibnamefont {Dyakonov}},\ and\
  \bibinfo {author} {\bibfnamefont {C.}~\bibnamefont {Deibel}},\ }\href@noop {}
  {\bibfield  {journal} {\bibinfo  {journal} {Physical Review B}\ }\textbf
  {\bibinfo {volume} {86}},\ \bibinfo {pages} {115302} (\bibinfo {year}
  {2012})}\BibitemShut {NoStop}%
\bibitem [{\citenamefont {Tvingstedt}\ and\ \citenamefont
  {Deibel}(2016)}]{tvingstedt2016temperature}%
  \BibitemOpen
  \bibfield  {author} {\bibinfo {author} {\bibfnamefont {K.}~\bibnamefont
  {Tvingstedt}}\ and\ \bibinfo {author} {\bibfnamefont {C.}~\bibnamefont
  {Deibel}},\ }\href@noop {} {\bibfield  {journal} {\bibinfo  {journal}
  {Advanced Energy Materials}\ }\textbf {\bibinfo {volume} {6}},\ \bibinfo
  {pages} {1502230} (\bibinfo {year} {2016})}\BibitemShut {NoStop}%
\bibitem [{\citenamefont {Perdig{\'o}n-Toro}\ \emph {et~al.}(2022)\citenamefont
  {Perdig{\'o}n-Toro}, \citenamefont {Phuong}, \citenamefont {Eller},
  \citenamefont {Freychet}, \citenamefont {Saglamkaya}, \citenamefont {Khan},
  \citenamefont {Wei}, \citenamefont {Zeiske}, \citenamefont {Kroh},
  \citenamefont {Wedler}, \citenamefont {K{\"o}hler}, \citenamefont {Armin},
  \citenamefont {Laquai}, \citenamefont {Herzig}, \citenamefont {Zou},
  \citenamefont {Shoaee},\ and\ \citenamefont
  {Neher}}]{perdigon2022understanding}%
  \BibitemOpen
  \bibfield  {author} {\bibinfo {author} {\bibfnamefont {L.}~\bibnamefont
  {Perdig{\'o}n-Toro}}, \bibinfo {author} {\bibfnamefont {L.~Q.}\ \bibnamefont
  {Phuong}}, \bibinfo {author} {\bibfnamefont {F.}~\bibnamefont {Eller}},
  \bibinfo {author} {\bibfnamefont {G.}~\bibnamefont {Freychet}}, \bibinfo
  {author} {\bibfnamefont {E.}~\bibnamefont {Saglamkaya}}, \bibinfo {author}
  {\bibfnamefont {J.~I.}\ \bibnamefont {Khan}}, \bibinfo {author}
  {\bibfnamefont {Q.}~\bibnamefont {Wei}}, \bibinfo {author} {\bibfnamefont
  {S.}~\bibnamefont {Zeiske}}, \bibinfo {author} {\bibfnamefont
  {D.}~\bibnamefont {Kroh}}, \bibinfo {author} {\bibfnamefont {S.}~\bibnamefont
  {Wedler}}, \bibinfo {author} {\bibfnamefont {A.}~\bibnamefont {K{\"o}hler}},
  \bibinfo {author} {\bibfnamefont {A.}~\bibnamefont {Armin}}, \bibinfo
  {author} {\bibfnamefont {F.}~\bibnamefont {Laquai}}, \bibinfo {author}
  {\bibfnamefont {E.~M.}\ \bibnamefont {Herzig}}, \bibinfo {author}
  {\bibfnamefont {Y.}~\bibnamefont {Zou}}, \bibinfo {author} {\bibfnamefont
  {S.}~\bibnamefont {Shoaee}},\ and\ \bibinfo {author} {\bibfnamefont
  {D.}~\bibnamefont {Neher}},\ }\href@noop {} {\bibfield  {journal} {\bibinfo
  {journal} {Advanced Energy Materials}\ }\textbf {\bibinfo {volume} {12}},\
  \bibinfo {pages} {2103422} (\bibinfo {year} {2022})}\BibitemShut {NoStop}%
\bibitem [{\citenamefont {MacKenzie}\ \emph {et~al.}(2011)\citenamefont
  {MacKenzie}, \citenamefont {Kirchartz}, \citenamefont {Dibb},\ and\
  \citenamefont {Nelson}}]{mackenzie2011modeling}%
  \BibitemOpen
  \bibfield  {author} {\bibinfo {author} {\bibfnamefont {R.~C.}\ \bibnamefont
  {MacKenzie}}, \bibinfo {author} {\bibfnamefont {T.}~\bibnamefont
  {Kirchartz}}, \bibinfo {author} {\bibfnamefont {G.~F.}\ \bibnamefont
  {Dibb}},\ and\ \bibinfo {author} {\bibfnamefont {J.}~\bibnamefont {Nelson}},\
  }\href@noop {} {\bibfield  {journal} {\bibinfo  {journal} {The Journal of
  Physical Chemistry C}\ }\textbf {\bibinfo {volume} {115}},\ \bibinfo {pages}
  {9806} (\bibinfo {year} {2011})}\BibitemShut {NoStop}%
\bibitem [{\citenamefont {MacKenzie}\ \emph {et~al.}(2012)\citenamefont
  {MacKenzie}, \citenamefont {Shuttle}, \citenamefont {Chabinyc},\ and\
  \citenamefont {Nelson}}]{mackenzie2012extracting}%
  \BibitemOpen
  \bibfield  {author} {\bibinfo {author} {\bibfnamefont {R.~C.}\ \bibnamefont
  {MacKenzie}}, \bibinfo {author} {\bibfnamefont {C.~G.}\ \bibnamefont
  {Shuttle}}, \bibinfo {author} {\bibfnamefont {M.~L.}\ \bibnamefont
  {Chabinyc}},\ and\ \bibinfo {author} {\bibfnamefont {J.}~\bibnamefont
  {Nelson}},\ }\href@noop {} {\bibfield  {journal} {\bibinfo  {journal}
  {Advanced Energy Materials}\ }\textbf {\bibinfo {volume} {2}},\ \bibinfo
  {pages} {662} (\bibinfo {year} {2012})}\BibitemShut {NoStop}%
\bibitem [{\citenamefont {Oelerich}\ \emph {et~al.}(2012)\citenamefont
  {Oelerich}, \citenamefont {Huemmer},\ and\ \citenamefont
  {Baranovskii}}]{oelerich2012find}%
  \BibitemOpen
  \bibfield  {author} {\bibinfo {author} {\bibfnamefont {J.}~\bibnamefont
  {Oelerich}}, \bibinfo {author} {\bibfnamefont {D.}~\bibnamefont {Huemmer}},\
  and\ \bibinfo {author} {\bibfnamefont {S.}~\bibnamefont {Baranovskii}},\
  }\href@noop {} {\bibfield  {journal} {\bibinfo  {journal} {Physical Review
  Letters}\ }\textbf {\bibinfo {volume} {108}},\ \bibinfo {pages} {226403}
  (\bibinfo {year} {2012})}\BibitemShut {NoStop}%
\bibitem [{\citenamefont {Noolandi}(1977)}]{noolandi1977multiple}%
  \BibitemOpen
  \bibfield  {author} {\bibinfo {author} {\bibfnamefont {J.}~\bibnamefont
  {Noolandi}},\ }\href@noop {} {\bibfield  {journal} {\bibinfo  {journal}
  {Physical Review B}\ }\textbf {\bibinfo {volume} {16}},\ \bibinfo {pages}
  {4466} (\bibinfo {year} {1977})}\BibitemShut {NoStop}%
\bibitem [{\citenamefont {Arkhipov}\ \emph {et~al.}(1984)\citenamefont
  {Arkhipov}, \citenamefont {Kolesnikov},\ and\ \citenamefont
  {Rudenko}}]{arkhipov1984dispersive}%
  \BibitemOpen
  \bibfield  {author} {\bibinfo {author} {\bibfnamefont {V.}~\bibnamefont
  {Arkhipov}}, \bibinfo {author} {\bibfnamefont {V.}~\bibnamefont
  {Kolesnikov}},\ and\ \bibinfo {author} {\bibfnamefont {A.}~\bibnamefont
  {Rudenko}},\ }\href@noop {} {\bibfield  {journal} {\bibinfo  {journal}
  {Journal of Physics D: Applied Physics}\ }\textbf {\bibinfo {volume} {17}},\
  \bibinfo {pages} {1241} (\bibinfo {year} {1984})}\BibitemShut {NoStop}%
\bibitem [{\citenamefont {Mark}\ and\ \citenamefont
  {Helfrich}(1962)}]{mark1962space}%
  \BibitemOpen
  \bibfield  {author} {\bibinfo {author} {\bibfnamefont {P.}~\bibnamefont
  {Mark}}\ and\ \bibinfo {author} {\bibfnamefont {W.}~\bibnamefont
  {Helfrich}},\ }\href@noop {} {\bibfield  {journal} {\bibinfo  {journal}
  {Journal of Applied Physics}\ }\textbf {\bibinfo {volume} {33}},\ \bibinfo
  {pages} {205} (\bibinfo {year} {1962})}\BibitemShut {NoStop}%
\bibitem [{\citenamefont {Paasch}\ and\ \citenamefont
  {Scheinert}(2010)}]{paasch2010charge}%
  \BibitemOpen
  \bibfield  {author} {\bibinfo {author} {\bibfnamefont {G.}~\bibnamefont
  {Paasch}}\ and\ \bibinfo {author} {\bibfnamefont {S.}~\bibnamefont
  {Scheinert}},\ }\href@noop {} {\bibfield  {journal} {\bibinfo  {journal}
  {Journal of Applied Physics}\ }\textbf {\bibinfo {volume} {107}},\ \bibinfo
  {pages} {104501} (\bibinfo {year} {2010})}\BibitemShut {NoStop}%
\bibitem [{\citenamefont {Garcia-Belmonte}(2010)}]{garcia2010temperature}%
  \BibitemOpen
  \bibfield  {author} {\bibinfo {author} {\bibfnamefont {G.}~\bibnamefont
  {Garcia-Belmonte}},\ }\href@noop {} {\bibfield  {journal} {\bibinfo
  {journal} {Solar Energy Materials and Solar Cells}\ }\textbf {\bibinfo
  {volume} {94}},\ \bibinfo {pages} {2166} (\bibinfo {year}
  {2010})}\BibitemShut {NoStop}%
\bibitem [{\citenamefont {Kirchartz}\ and\ \citenamefont
  {Nelson}(2012)}]{kirchartz2012meaning}%
  \BibitemOpen
  \bibfield  {author} {\bibinfo {author} {\bibfnamefont {T.}~\bibnamefont
  {Kirchartz}}\ and\ \bibinfo {author} {\bibfnamefont {J.}~\bibnamefont
  {Nelson}},\ }\href@noop {} {\bibfield  {journal} {\bibinfo  {journal}
  {Physical Review B}\ }\textbf {\bibinfo {volume} {86}},\ \bibinfo {pages}
  {165201} (\bibinfo {year} {2012})}\BibitemShut {NoStop}%
\bibitem [{\citenamefont {Burke}\ \emph {et~al.}(2015)\citenamefont {Burke},
  \citenamefont {Sweetnam}, \citenamefont {Vandewal},\ and\ \citenamefont
  {McGehee}}]{burke2015beyond}%
  \BibitemOpen
  \bibfield  {author} {\bibinfo {author} {\bibfnamefont {T.~M.}\ \bibnamefont
  {Burke}}, \bibinfo {author} {\bibfnamefont {S.}~\bibnamefont {Sweetnam}},
  \bibinfo {author} {\bibfnamefont {K.}~\bibnamefont {Vandewal}},\ and\
  \bibinfo {author} {\bibfnamefont {M.~D.}\ \bibnamefont {McGehee}},\
  }\href@noop {} {\bibfield  {journal} {\bibinfo  {journal} {Advanced Energy
  Materials}\ }\textbf {\bibinfo {volume} {5}},\ \bibinfo {pages} {1500123}
  (\bibinfo {year} {2015})}\BibitemShut {NoStop}%
\bibitem [{\citenamefont {Xiao}\ \emph {et~al.}(2020)\citenamefont {Xiao},
  \citenamefont {Calado}, \citenamefont {MacKenzie}, \citenamefont {Kirchartz},
  \citenamefont {Yan},\ and\ \citenamefont {Nelson}}]{xiao2020relationship}%
  \BibitemOpen
  \bibfield  {author} {\bibinfo {author} {\bibfnamefont {B.}~\bibnamefont
  {Xiao}}, \bibinfo {author} {\bibfnamefont {P.}~\bibnamefont {Calado}},
  \bibinfo {author} {\bibfnamefont {R.~C.}\ \bibnamefont {MacKenzie}}, \bibinfo
  {author} {\bibfnamefont {T.}~\bibnamefont {Kirchartz}}, \bibinfo {author}
  {\bibfnamefont {J.}~\bibnamefont {Yan}},\ and\ \bibinfo {author}
  {\bibfnamefont {J.}~\bibnamefont {Nelson}},\ }\href@noop {} {\bibfield
  {journal} {\bibinfo  {journal} {Physical Review Applied}\ }\textbf {\bibinfo
  {volume} {14}},\ \bibinfo {pages} {024034} (\bibinfo {year}
  {2020})}\BibitemShut {NoStop}%
\bibitem [{\citenamefont {Shockley}(1949)}]{shockley1949theory}%
  \BibitemOpen
  \bibfield  {author} {\bibinfo {author} {\bibfnamefont {W.}~\bibnamefont
  {Shockley}},\ }\href@noop {} {\bibfield  {journal} {\bibinfo  {journal} {Bell
  System Technical Journal}\ }\textbf {\bibinfo {volume} {28}},\ \bibinfo
  {pages} {435} (\bibinfo {year} {1949})}\BibitemShut {NoStop}%
\bibitem [{\citenamefont {W{\"u}rfel}\ \emph {et~al.}(2015)\citenamefont
  {W{\"u}rfel}, \citenamefont {Neher}, \citenamefont {Spies},\ and\
  \citenamefont {Albrecht}}]{wurfel2015impact}%
  \BibitemOpen
  \bibfield  {author} {\bibinfo {author} {\bibfnamefont {U.}~\bibnamefont
  {W{\"u}rfel}}, \bibinfo {author} {\bibfnamefont {D.}~\bibnamefont {Neher}},
  \bibinfo {author} {\bibfnamefont {A.}~\bibnamefont {Spies}},\ and\ \bibinfo
  {author} {\bibfnamefont {S.}~\bibnamefont {Albrecht}},\ }\href@noop {}
  {\bibfield  {journal} {\bibinfo  {journal} {Nature Communications}\ }\textbf
  {\bibinfo {volume} {6}},\ \bibinfo {pages} {6951} (\bibinfo {year}
  {2015})}\BibitemShut {NoStop}%
\bibitem [{\citenamefont {Noolandi}\ and\ \citenamefont
  {Hong}(1979)}]{noolandi1979theory}%
  \BibitemOpen
  \bibfield  {author} {\bibinfo {author} {\bibfnamefont {J.}~\bibnamefont
  {Noolandi}}\ and\ \bibinfo {author} {\bibfnamefont {K.}~\bibnamefont
  {Hong}},\ }\href@noop {} {\bibfield  {journal} {\bibinfo  {journal} {The
  Journal of Chemical Physics}\ }\textbf {\bibinfo {volume} {70}},\ \bibinfo
  {pages} {3230} (\bibinfo {year} {1979})}\BibitemShut {NoStop}%
\bibitem [{\citenamefont {Nikitenko}\ \emph {et~al.}(2001)\citenamefont
  {Nikitenko}, \citenamefont {Hertel},\ and\ \citenamefont
  {B{\"a}ssler}}]{nikitenko2001dispersive}%
  \BibitemOpen
  \bibfield  {author} {\bibinfo {author} {\bibfnamefont {V.}~\bibnamefont
  {Nikitenko}}, \bibinfo {author} {\bibfnamefont {D.}~\bibnamefont {Hertel}},\
  and\ \bibinfo {author} {\bibfnamefont {H.}~\bibnamefont {B{\"a}ssler}},\
  }\href@noop {} {\bibfield  {journal} {\bibinfo  {journal} {Chemical Physics
  Letters}\ }\textbf {\bibinfo {volume} {348}},\ \bibinfo {pages} {89}
  (\bibinfo {year} {2001})}\BibitemShut {NoStop}%
\bibitem [{\citenamefont {Hilczer}\ and\ \citenamefont
  {Tachiya}(2010)}]{hilczer2010unified}%
  \BibitemOpen
  \bibfield  {author} {\bibinfo {author} {\bibfnamefont {M.}~\bibnamefont
  {Hilczer}}\ and\ \bibinfo {author} {\bibfnamefont {M.}~\bibnamefont
  {Tachiya}},\ }\href@noop {} {\bibfield  {journal} {\bibinfo  {journal} {The
  Journal of Physical Chemistry C}\ }\textbf {\bibinfo {volume} {114}},\
  \bibinfo {pages} {6808} (\bibinfo {year} {2010})}\BibitemShut {NoStop}%
\bibitem [{\citenamefont {Tachiya}\ and\ \citenamefont
  {Seki}(2010)}]{tachiya2010theory}%
  \BibitemOpen
  \bibfield  {author} {\bibinfo {author} {\bibfnamefont {M.}~\bibnamefont
  {Tachiya}}\ and\ \bibinfo {author} {\bibfnamefont {K.}~\bibnamefont {Seki}},\
  }\href@noop {} {\bibfield  {journal} {\bibinfo  {journal} {Physical Review
  B}\ }\textbf {\bibinfo {volume} {82}},\ \bibinfo {pages} {085201} (\bibinfo
  {year} {2010})}\BibitemShut {NoStop}%
\bibitem [{\citenamefont {Gorenflot}\ \emph {et~al.}(2014)\citenamefont
  {Gorenflot}, \citenamefont {Heiber}, \citenamefont {Baumann}, \citenamefont
  {Lorrmann}, \citenamefont {Gunz}, \citenamefont {K{\"a}mpgen}, \citenamefont
  {Dyakonov},\ and\ \citenamefont {Deibel}}]{gorenflot2014nongeminate}%
  \BibitemOpen
  \bibfield  {author} {\bibinfo {author} {\bibfnamefont {J.}~\bibnamefont
  {Gorenflot}}, \bibinfo {author} {\bibfnamefont {M.~C.}\ \bibnamefont
  {Heiber}}, \bibinfo {author} {\bibfnamefont {A.}~\bibnamefont {Baumann}},
  \bibinfo {author} {\bibfnamefont {J.}~\bibnamefont {Lorrmann}}, \bibinfo
  {author} {\bibfnamefont {M.}~\bibnamefont {Gunz}}, \bibinfo {author}
  {\bibfnamefont {A.}~\bibnamefont {K{\"a}mpgen}}, \bibinfo {author}
  {\bibfnamefont {V.}~\bibnamefont {Dyakonov}},\ and\ \bibinfo {author}
  {\bibfnamefont {C.}~\bibnamefont {Deibel}},\ }\href@noop {} {\bibfield
  {journal} {\bibinfo  {journal} {Journal of Applied Physics}\ }\textbf
  {\bibinfo {volume} {115}},\ \bibinfo {pages} {144502} (\bibinfo {year}
  {2014})}\BibitemShut {NoStop}%
\bibitem [{\citenamefont {Langevin}(1903)}]{langevin1903recombinaison}%
  \BibitemOpen
  \bibfield  {author} {\bibinfo {author} {\bibfnamefont {P.}~\bibnamefont
  {Langevin}},\ }\href@noop {} {\bibfield  {journal} {\bibinfo  {journal}
  {Annales de Chimie et de Physique}\ }\textbf {\bibinfo {volume} {28}},\
  \bibinfo {pages} {433} (\bibinfo {year} {1903})}\BibitemShut {NoStop}%
\bibitem [{\citenamefont {Braun}(1984)}]{braun1984electric}%
  \BibitemOpen
  \bibfield  {author} {\bibinfo {author} {\bibfnamefont {C.~L.}\ \bibnamefont
  {Braun}},\ }\href@noop {} {\bibfield  {journal} {\bibinfo  {journal} {The
  Journal of Chemical Physics}\ }\textbf {\bibinfo {volume} {80}},\ \bibinfo
  {pages} {4157} (\bibinfo {year} {1984})}\BibitemShut {NoStop}%
\bibitem [{\citenamefont {Shoaee}\ \emph {et~al.}(2019)\citenamefont {Shoaee},
  \citenamefont {Armin}, \citenamefont {Stolterfoht}, \citenamefont {Hosseini},
  \citenamefont {Kurpiers},\ and\ \citenamefont {Neher}}]{shoaee2019decoding}%
  \BibitemOpen
  \bibfield  {author} {\bibinfo {author} {\bibfnamefont {S.}~\bibnamefont
  {Shoaee}}, \bibinfo {author} {\bibfnamefont {A.}~\bibnamefont {Armin}},
  \bibinfo {author} {\bibfnamefont {M.}~\bibnamefont {Stolterfoht}}, \bibinfo
  {author} {\bibfnamefont {S.~M.}\ \bibnamefont {Hosseini}}, \bibinfo {author}
  {\bibfnamefont {J.}~\bibnamefont {Kurpiers}},\ and\ \bibinfo {author}
  {\bibfnamefont {D.}~\bibnamefont {Neher}},\ }\href@noop {} {\bibfield
  {journal} {\bibinfo  {journal} {Solar RRL}\ }\textbf {\bibinfo {volume}
  {3}},\ \bibinfo {pages} {1900184} (\bibinfo {year} {2019})}\BibitemShut
  {NoStop}%
\bibitem [{\citenamefont {Koster}\ \emph {et~al.}(2006)\citenamefont {Koster},
  \citenamefont {Mihailetchi},\ and\ \citenamefont
  {Blom}}]{koster2006bimolecular}%
  \BibitemOpen
  \bibfield  {author} {\bibinfo {author} {\bibfnamefont {L.}~\bibnamefont
  {Koster}}, \bibinfo {author} {\bibfnamefont {V.}~\bibnamefont
  {Mihailetchi}},\ and\ \bibinfo {author} {\bibfnamefont {P.}~\bibnamefont
  {Blom}},\ }\href@noop {} {\bibfield  {journal} {\bibinfo  {journal} {Applied
  Physics Letters}\ }\textbf {\bibinfo {volume} {88}},\ \bibinfo {pages}
  {052104} (\bibinfo {year} {2006})}\BibitemShut {NoStop}%
\bibitem [{\citenamefont {Heiber}\ \emph {et~al.}(2015)\citenamefont {Heiber},
  \citenamefont {Baumbach}, \citenamefont {Dyakonov},\ and\ \citenamefont
  {Deibel}}]{heiber2015encounter}%
  \BibitemOpen
  \bibfield  {author} {\bibinfo {author} {\bibfnamefont {M.~C.}\ \bibnamefont
  {Heiber}}, \bibinfo {author} {\bibfnamefont {C.}~\bibnamefont {Baumbach}},
  \bibinfo {author} {\bibfnamefont {V.}~\bibnamefont {Dyakonov}},\ and\
  \bibinfo {author} {\bibfnamefont {C.}~\bibnamefont {Deibel}},\ }\href@noop {}
  {\bibfield  {journal} {\bibinfo  {journal} {Physical Review Letters}\
  }\textbf {\bibinfo {volume} {114}},\ \bibinfo {pages} {136602} (\bibinfo
  {year} {2015})}\BibitemShut {NoStop}%
\bibitem [{\citenamefont {Heiber}\ \emph {et~al.}(2016)\citenamefont {Heiber},
  \citenamefont {Nguyen},\ and\ \citenamefont {Deibel}}]{heiber2016charge}%
  \BibitemOpen
  \bibfield  {author} {\bibinfo {author} {\bibfnamefont {M.~C.}\ \bibnamefont
  {Heiber}}, \bibinfo {author} {\bibfnamefont {T.-Q.}\ \bibnamefont {Nguyen}},\
  and\ \bibinfo {author} {\bibfnamefont {C.}~\bibnamefont {Deibel}},\
  }\href@noop {} {\bibfield  {journal} {\bibinfo  {journal} {Physical Review
  B}\ }\textbf {\bibinfo {volume} {93}},\ \bibinfo {pages} {205204} (\bibinfo
  {year} {2016})}\BibitemShut {NoStop}%
\bibitem [{\citenamefont {Arkhipov}\ \emph {et~al.}(2001)\citenamefont
  {Arkhipov}, \citenamefont {Emelianova},\ and\ \citenamefont
  {Adriaenssens}}]{arkhipov2001effective}%
  \BibitemOpen
  \bibfield  {author} {\bibinfo {author} {\bibfnamefont {V.}~\bibnamefont
  {Arkhipov}}, \bibinfo {author} {\bibfnamefont {E.}~\bibnamefont
  {Emelianova}},\ and\ \bibinfo {author} {\bibfnamefont {G.}~\bibnamefont
  {Adriaenssens}},\ }\href@noop {} {\bibfield  {journal} {\bibinfo  {journal}
  {Physical Review B}\ }\textbf {\bibinfo {volume} {64}},\ \bibinfo {pages}
  {125125} (\bibinfo {year} {2001})}\BibitemShut {NoStop}%
\bibitem [{\citenamefont {Adriaenssens}\ \emph {et~al.}(1995)\citenamefont
  {Adriaenssens}, \citenamefont {Baranovskii}, \citenamefont {Fuhs},
  \citenamefont {Jansen},\ and\ \citenamefont
  {{\"O}kt{\"u}}}]{adriaenssens1995photoconductivity}%
  \BibitemOpen
  \bibfield  {author} {\bibinfo {author} {\bibfnamefont {G.}~\bibnamefont
  {Adriaenssens}}, \bibinfo {author} {\bibfnamefont {S.}~\bibnamefont
  {Baranovskii}}, \bibinfo {author} {\bibfnamefont {W.}~\bibnamefont {Fuhs}},
  \bibinfo {author} {\bibfnamefont {J.}~\bibnamefont {Jansen}},\ and\ \bibinfo
  {author} {\bibfnamefont {{\"O}.}~\bibnamefont {{\"O}kt{\"u}}},\ }\href@noop
  {} {\bibfield  {journal} {\bibinfo  {journal} {Physical Review B}\ }\textbf
  {\bibinfo {volume} {51}},\ \bibinfo {pages} {9661} (\bibinfo {year}
  {1995})}\BibitemShut {NoStop}%
\bibitem [{\citenamefont {Arkhipov}\ \emph {et~al.}(2006)\citenamefont
  {Arkhipov}, \citenamefont {Fishchuk}, \citenamefont {Kadashchuk},\ and\
  \citenamefont {B{\"a}ssler}}]{arkhipov2005charge}%
  \BibitemOpen
  \bibfield  {author} {\bibinfo {author} {\bibfnamefont {V.}~\bibnamefont
  {Arkhipov}}, \bibinfo {author} {\bibfnamefont {I.}~\bibnamefont {Fishchuk}},
  \bibinfo {author} {\bibfnamefont {A.}~\bibnamefont {Kadashchuk}},\ and\
  \bibinfo {author} {\bibfnamefont {H.}~\bibnamefont {B{\"a}ssler}},\ }\bibinfo
  {title} {Charge transport in disordered organic semiconductors},\ in\
  \href@noop {} {\emph {\bibinfo {booktitle} {Photophysics of Molecular
  Materials: From Single Molecules To Single Crystals}}}\ (\bibinfo
  {publisher} {Wiley-VCH Verlag GmbH \& Co.\ KGaA},\ \bibinfo {year} {2006})\
  Chap.~\bibinfo {chapter} {6}, pp.\ \bibinfo {pages} {261--366}\BibitemShut
  {NoStop}%
\bibitem [{\citenamefont {Van~Berkel}\ \emph {et~al.}(1993)\citenamefont
  {Van~Berkel}, \citenamefont {Powell}, \citenamefont {Franklin},\ and\
  \citenamefont {French}}]{van1993quality}%
  \BibitemOpen
  \bibfield  {author} {\bibinfo {author} {\bibfnamefont {C.}~\bibnamefont
  {Van~Berkel}}, \bibinfo {author} {\bibfnamefont {M.}~\bibnamefont {Powell}},
  \bibinfo {author} {\bibfnamefont {A.}~\bibnamefont {Franklin}},\ and\
  \bibinfo {author} {\bibfnamefont {I.}~\bibnamefont {French}},\ }\href@noop {}
  {\bibfield  {journal} {\bibinfo  {journal} {Journal of Applied Physics}\
  }\textbf {\bibinfo {volume} {73}},\ \bibinfo {pages} {5264} (\bibinfo {year}
  {1993})}\BibitemShut {NoStop}%
\bibitem [{\citenamefont {Kirchartz}\ \emph {et~al.}(2011)\citenamefont
  {Kirchartz}, \citenamefont {Pieters}, \citenamefont {Kirkpatrick},
  \citenamefont {Rau},\ and\ \citenamefont
  {Nelson}}]{kirchartz2011recombination}%
  \BibitemOpen
  \bibfield  {author} {\bibinfo {author} {\bibfnamefont {T.}~\bibnamefont
  {Kirchartz}}, \bibinfo {author} {\bibfnamefont {B.~E.}\ \bibnamefont
  {Pieters}}, \bibinfo {author} {\bibfnamefont {J.}~\bibnamefont
  {Kirkpatrick}}, \bibinfo {author} {\bibfnamefont {U.}~\bibnamefont {Rau}},\
  and\ \bibinfo {author} {\bibfnamefont {J.}~\bibnamefont {Nelson}},\
  }\href@noop {} {\bibfield  {journal} {\bibinfo  {journal} {Physical Review
  B}\ }\textbf {\bibinfo {volume} {83}},\ \bibinfo {pages} {115209} (\bibinfo
  {year} {2011})}\BibitemShut {NoStop}%
\bibitem [{\citenamefont {Pasveer}\ \emph {et~al.}(2005)\citenamefont
  {Pasveer}, \citenamefont {Cottaar}, \citenamefont {Tanase}, \citenamefont
  {Coehoorn}, \citenamefont {Bobbert}, \citenamefont {Blom}, \citenamefont
  {De~Leeuw},\ and\ \citenamefont {Michels}}]{pasveer2005unified}%
  \BibitemOpen
  \bibfield  {author} {\bibinfo {author} {\bibfnamefont {W.}~\bibnamefont
  {Pasveer}}, \bibinfo {author} {\bibfnamefont {J.}~\bibnamefont {Cottaar}},
  \bibinfo {author} {\bibfnamefont {C.}~\bibnamefont {Tanase}}, \bibinfo
  {author} {\bibfnamefont {R.}~\bibnamefont {Coehoorn}}, \bibinfo {author}
  {\bibfnamefont {P.}~\bibnamefont {Bobbert}}, \bibinfo {author} {\bibfnamefont
  {P.}~\bibnamefont {Blom}}, \bibinfo {author} {\bibfnamefont {D.}~\bibnamefont
  {De~Leeuw}},\ and\ \bibinfo {author} {\bibfnamefont {M.}~\bibnamefont
  {Michels}},\ }\href@noop {} {\bibfield  {journal} {\bibinfo  {journal}
  {Physical Review Letters}\ }\textbf {\bibinfo {volume} {94}},\ \bibinfo
  {pages} {206601} (\bibinfo {year} {2005})}\BibitemShut {NoStop}%
\bibitem [{\citenamefont {Wolf}\ and\ \citenamefont
  {Rauschenbach}(1963)}]{wolf1963series}%
  \BibitemOpen
  \bibfield  {author} {\bibinfo {author} {\bibfnamefont {M.}~\bibnamefont
  {Wolf}}\ and\ \bibinfo {author} {\bibfnamefont {H.}~\bibnamefont
  {Rauschenbach}},\ }\href@noop {} {\bibfield  {journal} {\bibinfo  {journal}
  {Advanced Energy Conversion}\ }\textbf {\bibinfo {volume} {3}},\ \bibinfo
  {pages} {455} (\bibinfo {year} {1963})}\BibitemShut {NoStop}%
\bibitem [{\citenamefont {Kirchartz}\ \emph {et~al.}(2013)\citenamefont
  {Kirchartz}, \citenamefont {Deledalle}, \citenamefont {Tuladhar},
  \citenamefont {Durrant},\ and\ \citenamefont
  {Nelson}}]{kirchartz2013differences}%
  \BibitemOpen
  \bibfield  {author} {\bibinfo {author} {\bibfnamefont {T.}~\bibnamefont
  {Kirchartz}}, \bibinfo {author} {\bibfnamefont {F.}~\bibnamefont
  {Deledalle}}, \bibinfo {author} {\bibfnamefont {P.~S.}\ \bibnamefont
  {Tuladhar}}, \bibinfo {author} {\bibfnamefont {J.~R.}\ \bibnamefont
  {Durrant}},\ and\ \bibinfo {author} {\bibfnamefont {J.}~\bibnamefont
  {Nelson}},\ }\href@noop {} {\bibfield  {journal} {\bibinfo  {journal} {The
  Journal of Physical Chemistry Letters}\ }\textbf {\bibinfo {volume} {4}},\
  \bibinfo {pages} {2371} (\bibinfo {year} {2013})}\BibitemShut {NoStop}%
\bibitem [{\citenamefont {W{\"o}pke}\ \emph {et~al.}(2022)\citenamefont
  {W{\"o}pke}, \citenamefont {G{\"o}hler}, \citenamefont {Saladina},
  \citenamefont {Du}, \citenamefont {Nian}, \citenamefont {Greve},
  \citenamefont {Zhu}, \citenamefont {Yallum}, \citenamefont {Hofstetter},
  \citenamefont {Becker-Koch}, \citenamefont {Li}, \citenamefont
  {Heum{\"u}ller}, \citenamefont {Milekhin}, \citenamefont {Zahn},
  \citenamefont {Brabec}, \citenamefont {Banerji}, \citenamefont {Vaynzof},
  \citenamefont {Herzig}, \citenamefont {MacKenzie},\ and\ \citenamefont
  {Deibel}}]{wopke2022traps}%
  \BibitemOpen
  \bibfield  {author} {\bibinfo {author} {\bibfnamefont {C.}~\bibnamefont
  {W{\"o}pke}}, \bibinfo {author} {\bibfnamefont {C.}~\bibnamefont
  {G{\"o}hler}}, \bibinfo {author} {\bibfnamefont {M.}~\bibnamefont
  {Saladina}}, \bibinfo {author} {\bibfnamefont {X.}~\bibnamefont {Du}},
  \bibinfo {author} {\bibfnamefont {L.}~\bibnamefont {Nian}}, \bibinfo {author}
  {\bibfnamefont {C.}~\bibnamefont {Greve}}, \bibinfo {author} {\bibfnamefont
  {C.}~\bibnamefont {Zhu}}, \bibinfo {author} {\bibfnamefont {K.~M.}\
  \bibnamefont {Yallum}}, \bibinfo {author} {\bibfnamefont {Y.~J.}\
  \bibnamefont {Hofstetter}}, \bibinfo {author} {\bibfnamefont
  {D.}~\bibnamefont {Becker-Koch}}, \bibinfo {author} {\bibfnamefont
  {N.}~\bibnamefont {Li}}, \bibinfo {author} {\bibfnamefont {T.}~\bibnamefont
  {Heum{\"u}ller}}, \bibinfo {author} {\bibfnamefont {I.}~\bibnamefont
  {Milekhin}}, \bibinfo {author} {\bibfnamefont {D.~R.~T.}\ \bibnamefont
  {Zahn}}, \bibinfo {author} {\bibfnamefont {C.~J.}\ \bibnamefont {Brabec}},
  \bibinfo {author} {\bibfnamefont {N.}~\bibnamefont {Banerji}}, \bibinfo
  {author} {\bibfnamefont {Y.}~\bibnamefont {Vaynzof}}, \bibinfo {author}
  {\bibfnamefont {E.~M.}\ \bibnamefont {Herzig}}, \bibinfo {author}
  {\bibfnamefont {R.~C.}\ \bibnamefont {MacKenzie}},\ and\ \bibinfo {author}
  {\bibfnamefont {C.}~\bibnamefont {Deibel}},\ }\href@noop {} {\bibfield
  {journal} {\bibinfo  {journal} {Nature Communications}\ }\textbf {\bibinfo
  {volume} {13}},\ \bibinfo {pages} {3786} (\bibinfo {year}
  {2022})}\BibitemShut {NoStop}%
\bibitem [{\citenamefont {Wehrspohn}\ \emph {et~al.}(2000)\citenamefont
  {Wehrspohn}, \citenamefont {Deane}, \citenamefont {French}, \citenamefont
  {Gale}, \citenamefont {Hewett}, \citenamefont {Powell},\ and\ \citenamefont
  {Robertson}}]{wehrspohn2000relative}%
  \BibitemOpen
  \bibfield  {author} {\bibinfo {author} {\bibfnamefont {R.}~\bibnamefont
  {Wehrspohn}}, \bibinfo {author} {\bibfnamefont {S.}~\bibnamefont {Deane}},
  \bibinfo {author} {\bibfnamefont {I.}~\bibnamefont {French}}, \bibinfo
  {author} {\bibfnamefont {I.}~\bibnamefont {Gale}}, \bibinfo {author}
  {\bibfnamefont {J.}~\bibnamefont {Hewett}}, \bibinfo {author} {\bibfnamefont
  {M.}~\bibnamefont {Powell}},\ and\ \bibinfo {author} {\bibfnamefont
  {J.}~\bibnamefont {Robertson}},\ }\href@noop {} {\bibfield  {journal}
  {\bibinfo  {journal} {Journal of Applied Physics}\ }\textbf {\bibinfo
  {volume} {87}},\ \bibinfo {pages} {144} (\bibinfo {year} {2000})}\BibitemShut
  {NoStop}%
\bibitem [{\citenamefont {Karki}\ \emph {et~al.}(2019)\citenamefont {Karki},
  \citenamefont {Vollbrecht}, \citenamefont {Dixon}, \citenamefont {Schopp},
  \citenamefont {Schrock}, \citenamefont {Reddy},\ and\ \citenamefont
  {Nguyen}}]{karki2019understanding}%
  \BibitemOpen
  \bibfield  {author} {\bibinfo {author} {\bibfnamefont {A.}~\bibnamefont
  {Karki}}, \bibinfo {author} {\bibfnamefont {J.}~\bibnamefont {Vollbrecht}},
  \bibinfo {author} {\bibfnamefont {A.~L.}\ \bibnamefont {Dixon}}, \bibinfo
  {author} {\bibfnamefont {N.}~\bibnamefont {Schopp}}, \bibinfo {author}
  {\bibfnamefont {M.}~\bibnamefont {Schrock}}, \bibinfo {author} {\bibfnamefont
  {G.~M.}\ \bibnamefont {Reddy}},\ and\ \bibinfo {author} {\bibfnamefont
  {T.-Q.}\ \bibnamefont {Nguyen}},\ }\href@noop {} {\bibfield  {journal}
  {\bibinfo  {journal} {Advanced Materials}\ }\textbf {\bibinfo {volume}
  {31}},\ \bibinfo {pages} {1903868} (\bibinfo {year} {2019})}\BibitemShut
  {NoStop}%
\bibitem [{\citenamefont {Wu}\ \emph {et~al.}(2020)\citenamefont {Wu},
  \citenamefont {Lee}, \citenamefont {Chin}, \citenamefont {Yao}, \citenamefont
  {Cha}, \citenamefont {Luke}, \citenamefont {Hou}, \citenamefont {Kim},\ and\
  \citenamefont {Durrant}}]{wu2020exceptionally}%
  \BibitemOpen
  \bibfield  {author} {\bibinfo {author} {\bibfnamefont {J.}~\bibnamefont
  {Wu}}, \bibinfo {author} {\bibfnamefont {J.}~\bibnamefont {Lee}}, \bibinfo
  {author} {\bibfnamefont {Y.-C.}\ \bibnamefont {Chin}}, \bibinfo {author}
  {\bibfnamefont {H.}~\bibnamefont {Yao}}, \bibinfo {author} {\bibfnamefont
  {H.}~\bibnamefont {Cha}}, \bibinfo {author} {\bibfnamefont {J.}~\bibnamefont
  {Luke}}, \bibinfo {author} {\bibfnamefont {J.}~\bibnamefont {Hou}}, \bibinfo
  {author} {\bibfnamefont {J.-S.}\ \bibnamefont {Kim}},\ and\ \bibinfo {author}
  {\bibfnamefont {J.~R.}\ \bibnamefont {Durrant}},\ }\href@noop {} {\bibfield
  {journal} {\bibinfo  {journal} {Energy \& Environmental Science}\ }\textbf
  {\bibinfo {volume} {13}},\ \bibinfo {pages} {2422} (\bibinfo {year}
  {2020})}\BibitemShut {NoStop}%
\bibitem [{\citenamefont {Yang}\ \emph {et~al.}(2020)\citenamefont {Yang},
  \citenamefont {Ma}, \citenamefont {Cheng}, \citenamefont {Xiao},
  \citenamefont {Luo}, \citenamefont {Chen}, \citenamefont {Luo}, \citenamefont
  {Liu}, \citenamefont {Lu},\ and\ \citenamefont {Yan}}]{yang2020compatible}%
  \BibitemOpen
  \bibfield  {author} {\bibinfo {author} {\bibfnamefont {T.}~\bibnamefont
  {Yang}}, \bibinfo {author} {\bibfnamefont {R.}~\bibnamefont {Ma}}, \bibinfo
  {author} {\bibfnamefont {H.}~\bibnamefont {Cheng}}, \bibinfo {author}
  {\bibfnamefont {Y.}~\bibnamefont {Xiao}}, \bibinfo {author} {\bibfnamefont
  {Z.}~\bibnamefont {Luo}}, \bibinfo {author} {\bibfnamefont {Y.}~\bibnamefont
  {Chen}}, \bibinfo {author} {\bibfnamefont {S.}~\bibnamefont {Luo}}, \bibinfo
  {author} {\bibfnamefont {T.}~\bibnamefont {Liu}}, \bibinfo {author}
  {\bibfnamefont {X.}~\bibnamefont {Lu}},\ and\ \bibinfo {author}
  {\bibfnamefont {H.}~\bibnamefont {Yan}},\ }\href@noop {} {\bibfield
  {journal} {\bibinfo  {journal} {Journal of Materials Chemistry A}\ }\textbf
  {\bibinfo {volume} {8}},\ \bibinfo {pages} {17706} (\bibinfo {year}
  {2020})}\BibitemShut {NoStop}%
\bibitem [{\citenamefont {Lange}\ \emph {et~al.}(2013)\citenamefont {Lange},
  \citenamefont {Kniepert}, \citenamefont {Pingel}, \citenamefont {Dumsch},
  \citenamefont {Allard}, \citenamefont {Janietz}, \citenamefont {Scherf},\
  and\ \citenamefont {Neher}}]{lange2013correlation}%
  \BibitemOpen
  \bibfield  {author} {\bibinfo {author} {\bibfnamefont {I.}~\bibnamefont
  {Lange}}, \bibinfo {author} {\bibfnamefont {J.}~\bibnamefont {Kniepert}},
  \bibinfo {author} {\bibfnamefont {P.}~\bibnamefont {Pingel}}, \bibinfo
  {author} {\bibfnamefont {I.}~\bibnamefont {Dumsch}}, \bibinfo {author}
  {\bibfnamefont {S.}~\bibnamefont {Allard}}, \bibinfo {author} {\bibfnamefont
  {S.}~\bibnamefont {Janietz}}, \bibinfo {author} {\bibfnamefont
  {U.}~\bibnamefont {Scherf}},\ and\ \bibinfo {author} {\bibfnamefont
  {D.}~\bibnamefont {Neher}},\ }\href@noop {} {\bibfield  {journal} {\bibinfo
  {journal} {The Journal of Physical Chemistry Letters}\ }\textbf {\bibinfo
  {volume} {4}},\ \bibinfo {pages} {3865} (\bibinfo {year} {2013})}\BibitemShut
  {NoStop}%
\bibitem [{\citenamefont {Collins}\ \emph {et~al.}(2016)\citenamefont
  {Collins}, \citenamefont {Proctor}, \citenamefont {Ran},\ and\ \citenamefont
  {Nguyen}}]{collins2016understanding}%
  \BibitemOpen
  \bibfield  {author} {\bibinfo {author} {\bibfnamefont {S.~D.}\ \bibnamefont
  {Collins}}, \bibinfo {author} {\bibfnamefont {C.~M.}\ \bibnamefont
  {Proctor}}, \bibinfo {author} {\bibfnamefont {N.~A.}\ \bibnamefont {Ran}},\
  and\ \bibinfo {author} {\bibfnamefont {T.-Q.}\ \bibnamefont {Nguyen}},\
  }\href@noop {} {\bibfield  {journal} {\bibinfo  {journal} {Advanced Energy
  Materials}\ }\textbf {\bibinfo {volume} {6}},\ \bibinfo {pages} {1501721}
  (\bibinfo {year} {2016})}\BibitemShut {NoStop}%
\bibitem [{\citenamefont {G{\"o}hler}\ and\ \citenamefont
  {Deibel}(2022)}]{goehler2022role}%
  \BibitemOpen
  \bibfield  {author} {\bibinfo {author} {\bibfnamefont {C.}~\bibnamefont
  {G{\"o}hler}}\ and\ \bibinfo {author} {\bibfnamefont {C.}~\bibnamefont
  {Deibel}},\ }\href@noop {} {\bibfield  {journal} {\bibinfo  {journal} {ACS
  Energy Letters}\ }\textbf {\bibinfo {volume} {7}},\ \bibinfo {pages} {2156}
  (\bibinfo {year} {2022})}\BibitemShut {NoStop}%
\end{thebibliography}%


\begin{thebibliography}{14}%
\makeatletter
\providecommand \@ifxundefined [1]{%
 \@ifx{#1\undefined}
}%
\providecommand \@ifnum [1]{%
 \ifnum #1\expandafter \@firstoftwo
 \else \expandafter \@secondoftwo
 \fi
}%
\providecommand \@ifx [1]{%
 \ifx #1\expandafter \@firstoftwo
 \else \expandafter \@secondoftwo
 \fi
}%
\providecommand \natexlab [1]{#1}%
\providecommand \enquote  [1]{``#1''}%
\providecommand \bibnamefont  [1]{#1}%
\providecommand \bibfnamefont [1]{#1}%
\providecommand \citenamefont [1]{#1}%
\providecommand \href@noop [0]{\@secondoftwo}%
\providecommand \href [0]{\begingroup \@sanitize@url \@href}%
\providecommand \@href[1]{\@@startlink{#1}\@@href}%
\providecommand \@@href[1]{\endgroup#1\@@endlink}%
\providecommand \@sanitize@url [0]{\catcode `\\12\catcode `\$12\catcode
  `\&12\catcode `\#12\catcode `\^12\catcode `\_12\catcode `\%12\relax}%
\providecommand \@@startlink[1]{}%
\providecommand \@@endlink[0]{}%
\providecommand \url  [0]{\begingroup\@sanitize@url \@url }%
\providecommand \@url [1]{\endgroup\@href {#1}{\urlprefix }}%
\providecommand \urlprefix  [0]{URL }%
\providecommand \Eprint [0]{\href }%
\providecommand \doibase [0]{https://doi.org/}%
\providecommand \selectlanguage [0]{\@gobble}%
\providecommand \bibinfo  [0]{\@secondoftwo}%
\providecommand \bibfield  [0]{\@secondoftwo}%
\providecommand \translation [1]{[#1]}%
\providecommand \BibitemOpen [0]{}%
\providecommand \bibitemStop [0]{}%
\providecommand \bibitemNoStop [0]{.\EOS\space}%
\providecommand \EOS [0]{\spacefactor3000\relax}%
\providecommand \BibitemShut  [1]{\csname bibitem#1\endcsname}%
\let\auto@bib@innerbib\@empty
\bibitem [{\citenamefont {Fitzner}\ \emph {et~al.}(2012)\citenamefont
  {Fitzner}, \citenamefont {Mena-Osteritz}, \citenamefont {Mishra},
  \citenamefont {Schulz}, \citenamefont {Reinold}, \citenamefont {Weil},
  \citenamefont {K{\"o}rner}, \citenamefont {Ziehlke}, \citenamefont
  {Elschner}, \citenamefont {Leo}, \citenamefont {Riede}, \citenamefont
  {Pfeiffer}, \citenamefont {Uhrich},\ and\ \citenamefont
  {B{\"a}uerle}}]{Fitzner2012}%
  \BibitemOpen
  \bibfield  {author} {\bibinfo {author} {\bibfnamefont {R.}~\bibnamefont
  {Fitzner}}, \bibinfo {author} {\bibfnamefont {E.}~\bibnamefont
  {Mena-Osteritz}}, \bibinfo {author} {\bibfnamefont {A.}~\bibnamefont
  {Mishra}}, \bibinfo {author} {\bibfnamefont {G.}~\bibnamefont {Schulz}},
  \bibinfo {author} {\bibfnamefont {E.}~\bibnamefont {Reinold}}, \bibinfo
  {author} {\bibfnamefont {M.}~\bibnamefont {Weil}}, \bibinfo {author}
  {\bibfnamefont {C.}~\bibnamefont {K{\"o}rner}}, \bibinfo {author}
  {\bibfnamefont {H.}~\bibnamefont {Ziehlke}}, \bibinfo {author} {\bibfnamefont
  {C.}~\bibnamefont {Elschner}}, \bibinfo {author} {\bibfnamefont
  {K.}~\bibnamefont {Leo}}, \bibinfo {author} {\bibfnamefont {M.}~\bibnamefont
  {Riede}}, \bibinfo {author} {\bibfnamefont {M.}~\bibnamefont {Pfeiffer}},
  \bibinfo {author} {\bibfnamefont {C.}~\bibnamefont {Uhrich}},\ and\ \bibinfo
  {author} {\bibfnamefont {P.}~\bibnamefont {B{\"a}uerle}},\ }\href@noop {}
  {\bibfield  {journal} {\bibinfo  {journal} {Journal of the American Chemical
  Society}\ }\textbf {\bibinfo {volume} {134}},\ \bibinfo {pages} {11064}
  (\bibinfo {year} {2012})}\BibitemShut {NoStop}%
\bibitem [{\citenamefont {Mark}\ and\ \citenamefont
  {Helfrich}(1962)}]{mark1962space}%
  \BibitemOpen
  \bibfield  {author} {\bibinfo {author} {\bibfnamefont {P.}~\bibnamefont
  {Mark}}\ and\ \bibinfo {author} {\bibfnamefont {W.}~\bibnamefont
  {Helfrich}},\ }\href@noop {} {\bibfield  {journal} {\bibinfo  {journal}
  {Journal of Applied Physics}\ }\textbf {\bibinfo {volume} {33}},\ \bibinfo
  {pages} {205} (\bibinfo {year} {1962})}\BibitemShut {NoStop}%
\bibitem [{\citenamefont {Blakesley}\ and\ \citenamefont
  {Neher}(2011)}]{blakesley2011relationship}%
  \BibitemOpen
  \bibfield  {author} {\bibinfo {author} {\bibfnamefont {J.~C.}\ \bibnamefont
  {Blakesley}}\ and\ \bibinfo {author} {\bibfnamefont {D.}~\bibnamefont
  {Neher}},\ }\href@noop {} {\bibfield  {journal} {\bibinfo  {journal}
  {Physical Review B}\ }\textbf {\bibinfo {volume} {84}},\ \bibinfo {pages}
  {075210} (\bibinfo {year} {2011})}\BibitemShut {NoStop}%
\bibitem [{\citenamefont {Kirchartz}\ and\ \citenamefont
  {Nelson}(2012)}]{kirchartz2012meaning}%
  \BibitemOpen
  \bibfield  {author} {\bibinfo {author} {\bibfnamefont {T.}~\bibnamefont
  {Kirchartz}}\ and\ \bibinfo {author} {\bibfnamefont {J.}~\bibnamefont
  {Nelson}},\ }\href@noop {} {\bibfield  {journal} {\bibinfo  {journal}
  {Physical Review B}\ }\textbf {\bibinfo {volume} {86}},\ \bibinfo {pages}
  {165201} (\bibinfo {year} {2012})}\BibitemShut {NoStop}%
\bibitem [{\citenamefont {Hofacker}\ and\ \citenamefont
  {Neher}(2017)}]{hofacker2017dispersive}%
  \BibitemOpen
  \bibfield  {author} {\bibinfo {author} {\bibfnamefont {A.}~\bibnamefont
  {Hofacker}}\ and\ \bibinfo {author} {\bibfnamefont {D.}~\bibnamefont
  {Neher}},\ }\href@noop {} {\bibfield  {journal} {\bibinfo  {journal}
  {Physical Review B}\ }\textbf {\bibinfo {volume} {96}},\ \bibinfo {pages}
  {245204} (\bibinfo {year} {2017})}\BibitemShut {NoStop}%
\bibitem [{\citenamefont {Xiao}\ \emph {et~al.}(2020)\citenamefont {Xiao},
  \citenamefont {Calado}, \citenamefont {MacKenzie}, \citenamefont {Kirchartz},
  \citenamefont {Yan},\ and\ \citenamefont {Nelson}}]{xiao2020relationship}%
  \BibitemOpen
  \bibfield  {author} {\bibinfo {author} {\bibfnamefont {B.}~\bibnamefont
  {Xiao}}, \bibinfo {author} {\bibfnamefont {P.}~\bibnamefont {Calado}},
  \bibinfo {author} {\bibfnamefont {R.~C.}\ \bibnamefont {MacKenzie}}, \bibinfo
  {author} {\bibfnamefont {T.}~\bibnamefont {Kirchartz}}, \bibinfo {author}
  {\bibfnamefont {J.}~\bibnamefont {Yan}},\ and\ \bibinfo {author}
  {\bibfnamefont {J.}~\bibnamefont {Nelson}},\ }\href@noop {} {\bibfield
  {journal} {\bibinfo  {journal} {Physical Review Applied}\ }\textbf {\bibinfo
  {volume} {14}},\ \bibinfo {pages} {024034} (\bibinfo {year}
  {2020})}\BibitemShut {NoStop}%
\bibitem [{\citenamefont {Arkhipov}\ \emph {et~al.}(1984)\citenamefont
  {Arkhipov}, \citenamefont {Kolesnikov},\ and\ \citenamefont
  {Rudenko}}]{arkhipov1984dispersive}%
  \BibitemOpen
  \bibfield  {author} {\bibinfo {author} {\bibfnamefont {V.}~\bibnamefont
  {Arkhipov}}, \bibinfo {author} {\bibfnamefont {V.}~\bibnamefont
  {Kolesnikov}},\ and\ \bibinfo {author} {\bibfnamefont {A.}~\bibnamefont
  {Rudenko}},\ }\href@noop {} {\bibfield  {journal} {\bibinfo  {journal}
  {Journal of Physics D: Applied Physics}\ }\textbf {\bibinfo {volume} {17}},\
  \bibinfo {pages} {1241} (\bibinfo {year} {1984})}\BibitemShut {NoStop}%
\bibitem [{\citenamefont {Paasch}\ and\ \citenamefont
  {Scheinert}(2010)}]{paasch2010charge}%
  \BibitemOpen
  \bibfield  {author} {\bibinfo {author} {\bibfnamefont {G.}~\bibnamefont
  {Paasch}}\ and\ \bibinfo {author} {\bibfnamefont {S.}~\bibnamefont
  {Scheinert}},\ }\href@noop {} {\bibfield  {journal} {\bibinfo  {journal}
  {Journal of Applied Physics}\ }\textbf {\bibinfo {volume} {107}},\ \bibinfo
  {pages} {104501} (\bibinfo {year} {2010})}\BibitemShut {NoStop}%
\bibitem [{\citenamefont {Baranovskii}(2014)}]{baranovskii2014theoretical}%
  \BibitemOpen
  \bibfield  {author} {\bibinfo {author} {\bibfnamefont {S.}~\bibnamefont
  {Baranovskii}},\ }\href@noop {} {\bibfield  {journal} {\bibinfo  {journal}
  {Physica Status Solidi (b)}\ }\textbf {\bibinfo {volume} {251}},\ \bibinfo
  {pages} {487} (\bibinfo {year} {2014})}\BibitemShut {NoStop}%
\bibitem [{\citenamefont {Kirchartz}\ \emph {et~al.}(2011)\citenamefont
  {Kirchartz}, \citenamefont {Pieters}, \citenamefont {Kirkpatrick},
  \citenamefont {Rau},\ and\ \citenamefont
  {Nelson}}]{kirchartz2011recombination}%
  \BibitemOpen
  \bibfield  {author} {\bibinfo {author} {\bibfnamefont {T.}~\bibnamefont
  {Kirchartz}}, \bibinfo {author} {\bibfnamefont {B.~E.}\ \bibnamefont
  {Pieters}}, \bibinfo {author} {\bibfnamefont {J.}~\bibnamefont
  {Kirkpatrick}}, \bibinfo {author} {\bibfnamefont {U.}~\bibnamefont {Rau}},\
  and\ \bibinfo {author} {\bibfnamefont {J.}~\bibnamefont {Nelson}},\
  }\href@noop {} {\bibfield  {journal} {\bibinfo  {journal} {Physical Review
  B}\ }\textbf {\bibinfo {volume} {83}},\ \bibinfo {pages} {115209} (\bibinfo
  {year} {2011})}\BibitemShut {NoStop}%
\bibitem [{\citenamefont {Lange}\ \emph {et~al.}(2013)\citenamefont {Lange},
  \citenamefont {Kniepert}, \citenamefont {Pingel}, \citenamefont {Dumsch},
  \citenamefont {Allard}, \citenamefont {Janietz}, \citenamefont {Scherf},\
  and\ \citenamefont {Neher}}]{lange2013correlation}%
  \BibitemOpen
  \bibfield  {author} {\bibinfo {author} {\bibfnamefont {I.}~\bibnamefont
  {Lange}}, \bibinfo {author} {\bibfnamefont {J.}~\bibnamefont {Kniepert}},
  \bibinfo {author} {\bibfnamefont {P.}~\bibnamefont {Pingel}}, \bibinfo
  {author} {\bibfnamefont {I.}~\bibnamefont {Dumsch}}, \bibinfo {author}
  {\bibfnamefont {S.}~\bibnamefont {Allard}}, \bibinfo {author} {\bibfnamefont
  {S.}~\bibnamefont {Janietz}}, \bibinfo {author} {\bibfnamefont
  {U.}~\bibnamefont {Scherf}},\ and\ \bibinfo {author} {\bibfnamefont
  {D.}~\bibnamefont {Neher}},\ }\href@noop {} {\bibfield  {journal} {\bibinfo
  {journal} {The Journal of Physical Chemistry Letters}\ }\textbf {\bibinfo
  {volume} {4}},\ \bibinfo {pages} {3865} (\bibinfo {year} {2013})}\BibitemShut
  {NoStop}%
\bibitem [{\citenamefont {Collins}\ \emph {et~al.}(2016)\citenamefont
  {Collins}, \citenamefont {Proctor}, \citenamefont {Ran},\ and\ \citenamefont
  {Nguyen}}]{collins2016understanding}%
  \BibitemOpen
  \bibfield  {author} {\bibinfo {author} {\bibfnamefont {S.~D.}\ \bibnamefont
  {Collins}}, \bibinfo {author} {\bibfnamefont {C.~M.}\ \bibnamefont
  {Proctor}}, \bibinfo {author} {\bibfnamefont {N.~A.}\ \bibnamefont {Ran}},\
  and\ \bibinfo {author} {\bibfnamefont {T.-Q.}\ \bibnamefont {Nguyen}},\
  }\href@noop {} {\bibfield  {journal} {\bibinfo  {journal} {Advanced Energy
  Materials}\ }\textbf {\bibinfo {volume} {6}},\ \bibinfo {pages} {1501721}
  (\bibinfo {year} {2016})}\BibitemShut {NoStop}%
\bibitem [{\citenamefont {Garcia-Belmonte}(2010)}]{garcia2010temperature}%
  \BibitemOpen
  \bibfield  {author} {\bibinfo {author} {\bibfnamefont {G.}~\bibnamefont
  {Garcia-Belmonte}},\ }\href@noop {} {\bibfield  {journal} {\bibinfo
  {journal} {Solar Energy Materials and Solar Cells}\ }\textbf {\bibinfo
  {volume} {94}},\ \bibinfo {pages} {2166} (\bibinfo {year}
  {2010})}\BibitemShut {NoStop}%
\bibitem [{\citenamefont {Burke}\ \emph {et~al.}(2015)\citenamefont {Burke},
  \citenamefont {Sweetnam}, \citenamefont {Vandewal},\ and\ \citenamefont
  {McGehee}}]{burke2015beyond}%
  \BibitemOpen
  \bibfield  {author} {\bibinfo {author} {\bibfnamefont {T.~M.}\ \bibnamefont
  {Burke}}, \bibinfo {author} {\bibfnamefont {S.}~\bibnamefont {Sweetnam}},
  \bibinfo {author} {\bibfnamefont {K.}~\bibnamefont {Vandewal}},\ and\
  \bibinfo {author} {\bibfnamefont {M.~D.}\ \bibnamefont {McGehee}},\
  }\href@noop {} {\bibfield  {journal} {\bibinfo  {journal} {Advanced Energy
  Materials}\ }\textbf {\bibinfo {volume} {5}},\ \bibinfo {pages} {1500123}
  (\bibinfo {year} {2015})}\BibitemShut {NoStop}%
\end{thebibliography}%

\end{document}


\title{Power-law density of states in organic solar cells revealed by the open-circuit voltage dependence of the ideality factor}

\author{Maria Saladina}
\affiliation{Institut für Physik, Technische Universität Chemnitz, 09126 Chemnitz, Germany}

\author{Christopher Wöpke}
\affiliation{Institut für Physik, Technische Universität Chemnitz, 09126 Chemnitz, Germany}

\author{Clemens Göhler}
\affiliation{Institut für Physik, Technische Universität Chemnitz, 09126 Chemnitz, Germany}

\author{Ivan Ramirez}
\affiliation{Heliatek GmbH, 01139 Dresden, Germany}

\author{Olga Gerdes}
\affiliation{Heliatek GmbH, 01139 Dresden, Germany}

\author{Chao Liu}
\affiliation{Institute of Materials for Electronics and Energy Technology (i-MEET), Friedrich-Alexander-Universität Erlangen-Nürnberg, 91054 Erlangen, Germany}
\affiliation{Helmholtz Institute Erlangen-Nürnberg for Renewable Energy (HI ERN), 91058 Erlangen, Germany}

\author{Ning Li}
\affiliation{Institute of Materials for Electronics and Energy Technology (i-MEET), Friedrich-Alexander-Universität Erlangen-Nürnberg, 91054 Erlangen, Germany}
\affiliation{Helmholtz Institute Erlangen-Nürnberg for Renewable Energy (HI ERN), 91058 Erlangen, Germany}
\affiliation{State Key Laboratory of Luminescent Materials and Devices, Institute of Polymer Optoelectronic Materials and Devices, School of Materials Science and Engineering, South China University of Technology, 510640 Guangzhou, China}

\author{Thomas Heumüller}
\affiliation{Institute of Materials for Electronics and Energy Technology (i-MEET), Friedrich-Alexander-Universität Erlangen-Nürnberg, 91054 Erlangen, Germany}
\affiliation{Helmholtz Institute Erlangen-Nürnberg for Renewable Energy (HI ERN), 91058 Erlangen, Germany}

\author{Christoph J. Brabec}
\affiliation{Institute of Materials for Electronics and Energy Technology (i-MEET), Friedrich-Alexander-Universität Erlangen-Nürnberg, 91054 Erlangen, Germany}
\affiliation{Helmholtz Institute Erlangen-Nürnberg for Renewable Energy (HI ERN), 91058 Erlangen, Germany}

\author{Karsten Walzer}
\affiliation{Heliatek GmbH, 01139 Dresden, Germany}

\author{Martin Pfeiffer}
\affiliation{Heliatek GmbH, 01139 Dresden, Germany}

\author{Carsten Deibel}
\affiliation{Institut für Physik, Technische Universität Chemnitz, 09126 Chemnitz, Germany}

\begin{center}
    \large
    Supplemental material
    \vspace{0.2cm}
\end{center}

\maketitle

\section{Experimental methods}\label{sec:S1}

\subsection{Materials}

\begin{figure}[!htb]\centering
    \includegraphics[width=0.4\textwidth]{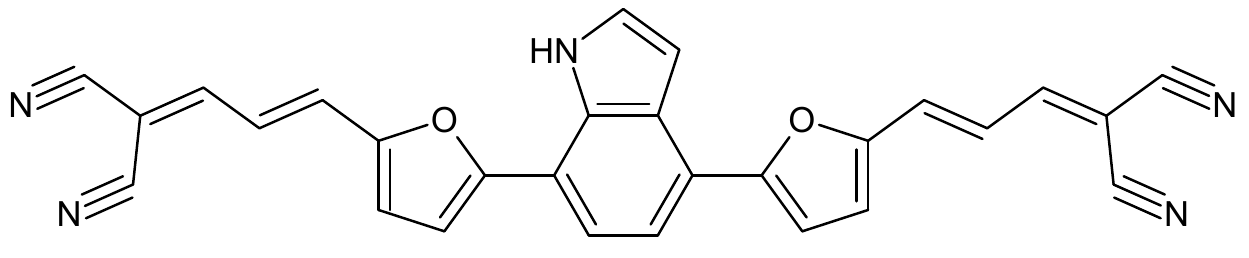}
    \caption{Molecular structure of DCV-V-Fu-Ind-Fu-V. }
\label{fig:S01}\end{figure}

\textbf{DCV-V-Fu-Ind-Fu-V:C$_{60}$}. DCV-V-Fu-Ind-Fu-V is a proprietary, oligomeric absorber material of Heliatek GmbH as shown in Figure~\ref{fig:S01}. It belongs to the the class of A-D-A (acceptor - donor - acceptor) type molecules, similar to the well-known DCV-oligothiophenes like DCV5T \cite{Fitzner2012}. A mixture with Fullerene C$_\text{60}$ is easily achieved by thermal co-evaporation in ultrahigh vacuum. 

\begin{figure}[!htb]\centering
    \includegraphics[width=0.4\textwidth]{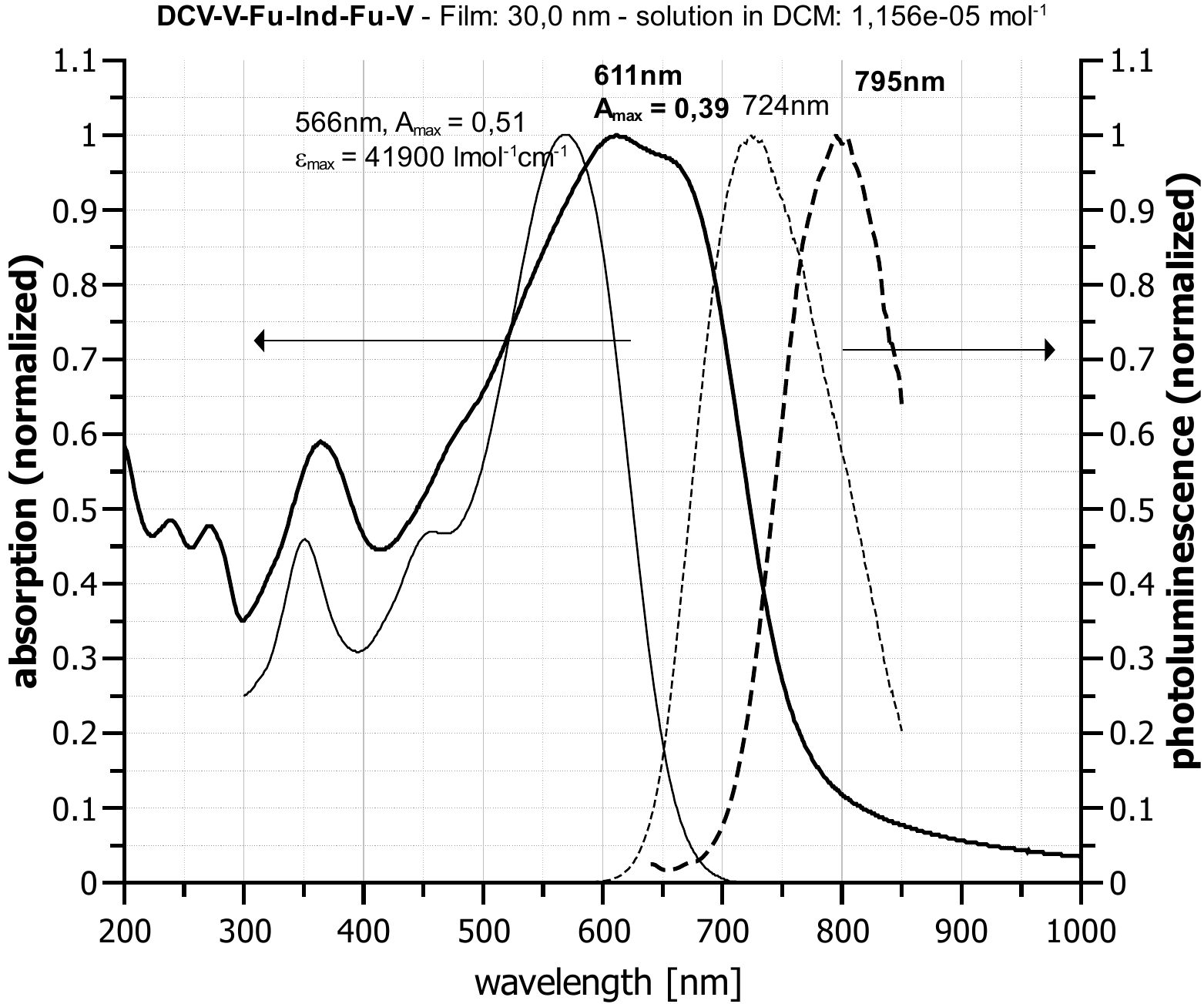}
    \caption{Optical characteristics of DCV-V-Fu-Ind-Fu-V in a 30 nm thick film (bold lines) and in dicyanomethane solution (thin lines). Straight lines refer to absorption, dashed lines represent photoluminescence data. }
\label{fig:S02}\end{figure}

\subsection{Device fabrication} 
\textbf{a-Si:H}. Hydrogenated amorphous silicon solar cell was purchased from Lemo-Solar GmbH. 
\vspace{0.3cm}

\textbf{P3HT:PC$_{61}$BM}. P3HT and PC$_\text{61}$BM were purchased from Ossila, and used without further purification. The active layer blend solution (1:0.8 w/w, 30\,mg\,ml$^{-1}$ in chlorobenzene) was stirred overnight at 50\,$^{\circ}$C in the glovebox. A glass substrate with pre-patterned indium tin oxide (ITO) was cleaned in ultrasonic bath with detergent, acetone, isopropanol and deionised water. After exposure to low-pressure oxygen plasma for 5\,min, a 35\,nm layer of poly(3,4-ethylenedioxythiophene) polystyrene sulfonate (PEDOT:PSS, Clevios AI~4083) was spin-coated and annealed at 140\,$^{\circ}$C for 10\,min. The substrate was transferred to a nitrogen-filled glovebox, where the blend solution was spin-coated to yield an active layer thickness of ca.\ 130\,nm, which was annealed at 150\,$^{\circ}$C for 10\,min. The device was completed by 2\,nm of thermally evaporated Ca and 150\,nm of Al with a base pressure below 10$^{-6}$\,mbar through a shadow mask, and had an active area of 4.0\,mm$^2$. 
\vspace{0.3cm}

\textbf{PM6:Y6}.  
PM6 was purchased from Solarmer Materials Inc (Beijing, China) and Y6 from Derthon Optoelectronic Materials Science Technology Co LTD (Shenzhen, China) and used as received. 
The active layer blend solution (1:1.2 w/w, 22\,mg\,ml$^{-1}$ in chloroform with 0.5 vol.-$\%$ of 1-chloronaphthalene) was stirred overnight at room temperature. ITO-coated glass substrate was cleaned with deionised water and isopropanol. The ZnO nanoparticle (N10, Avantama AG) solution was spin-coated onto ITO substrate at 3000 rpm for 30\,s and thermally annealed at 200\,$^{\circ}$C for 30 min. The substrate was then transferred into a nitrogen-filled glovebox. The blend solution was spin-coated resulting in a 200\,nm thick active layer and thermally annealed at 100\,$^{\circ}$C for 10\,min. The solar cell was completed by 10\,nm of MoO$_\text{x}$ and 100\,nm of Ag, thermally evaporated through a shadow mask, yielding an active area of $1.04~\text{cm}^2$. 
\vspace{0.3cm}

\textbf{DCV-V-Fu-Ind-Fu-V:C$_{60}$}. 
All the layers were deposited through thermal evaporation in an ultra-high vacuum chamber onto a cleaned glass substrate with structured ITO. An electron transport layer consisted of 5\,nm of n-doped (20\,\%) C$_\text{60}$ and 15\,nm of C$_\text{60}$. The active layer composed of DCV-V-Fu-Ind-Fu-V:C$_\text{60}$ (2:1 w/w) was co-evaporated onto a 50\,$^{\circ}$C hot substrate, which was subsequently cooled down for 60\,min. After that, the substrate was covered with a hole transport layer consisting of 10\,nm of BPAPF (Lumtec, Taiwan) and 20\,nm of p-doped (10\,\%) BPAPF. The solar cell was finished by an 100 nm thick Al electrode, and had an active area of 6.44\,mm$^2$. After fabrication, the device was encapsulated under nitrogen atmosphere with a glass lid. 
\vspace{0.3cm}

\subsection{Suns--$V_\mathrm{oc}$ measurements}
Samples were excited by an Omicron~LDM~A350 continuous wave laser at a wavelength of 515\,nm. The illumination intensity was varied by Standa motorised filter wheels with Thorlabs neutral density filters. Open-circuit voltage was measured with a Keithley~2634b source measure unit. 
During the experiment the sample was held in a Linkam Scientific~LTS420 cryostat which maintained low temperature by constant flow of liquid nitrogen using Linkam Scientific~LNP96-S liquid nitrogen pump and Linkam Scientific~T96-S temperature controller. 

\section{Derivation of the recombination rate for different DOS distributions}\label{sec:S2}
To make the link between the energetic picture and the ideality factor in the diode equation, the charge carrier density in Eq.\,(3) is expressed in terms of the effective band gap $E_\mathrm{g}$ and the open-circuit voltage $V_\mathrm{oc}$. This is achieved by relating the total densities of electrons and holes and then expressing $V_\mathrm{oc}$ through the quasi-Fermi level splitting. The resulting expression for the charge carrier concentration $n$, along with the relevant equation for $\theta$, replace the terms in Eq.\,(3).

\subsection{Charge carrier statistics}

The total charge carrier density in the exponential DOS depends on the characteristic distribution width $E_\mathrm{U}$ as~\cite{mark1962space,blakesley2011relationship,kirchartz2012meaning,hofacker2017dispersive,xiao2020relationship}
\begin{equation}\label{eq:S02}
    n_{\text{exp}} = N_0 \cdot \exp{\left( \frac{E_F}{E_\mathrm{U}}\right)} , 
\end{equation}
where $N_0$ stands for the total density of states and $E_F$ is the quasi-Fermi level relative to the transport energy. The transport energy is defined so that the probability of charge carrier release from a trap state with this energy is approaching unity~\cite{arkhipov1984dispersive}. 
The trapping factor in the exponential DOS depends on the quasi-Fermi level position and can be related to the charge carrier density in Eq.\,\eqref{eq:S02}~\cite{mark1962space,arkhipov1984dispersive,hofacker2017dispersive}
\begin{equation}\begin{split}\label{eq:S03}
    \theta_\text{exp} &= \exp{\left(\frac{E_F}{k_BT}\right)} \cdot \exp{\left(-\frac{E_F}{E_\mathrm{U}}\right)} \\
    &= N_0^{1-\lambda} \cdot n_\text{exp}^{\lambda-1} .  
\end{split}\end{equation}
Here the dispersion parameter $\lambda$ is defined as $\lambda = E_\mathrm{U}/k_BT$ with the Boltzmann constant $k_B$ and temperature $T$.
\vspace{0.3cm}

The concentration of charge carriers in the gaussian DOS is given by~\cite{paasch2010charge,blakesley2011relationship,hofacker2017dispersive}
\begin{equation}\label{eq:S05}
    n_{\text{gsn}} = N_0 \cdot \exp{\left(\frac{\sigma^2}{2(k_BT)^2}\right)} \exp{\left( \frac{E_F}{k_BT}\right)} , 
\end{equation}
where the disorder parameter $\sigma$ also relates the mobile charge carrier density $n_c$ to the total $n$, resulting in the energy-independent trapping factor~\cite{arkhipov1984dispersive,blakesley2011relationship,baranovskii2014theoretical,hofacker2017dispersive}
\begin{equation}\label{eq:S06}
    \theta_\text{gsn} = \exp{\left(-\frac{\sigma^2}{2(k_BT)^2}\right)} . 
\end{equation}

\subsubsection{Exponential DOS} 

If both charge carrier types occupy exponential DOS, the densities of electrons $n$ and holes $p$ are defined by Eq.\,\eqref{eq:S02}. 
With the relation $n = p$, and, assuming that $E_\mathrm{U}$ is the same for electrons and holes, the open-circuit voltage $V_\mathrm{oc}$ derived via the quasi-Fermi level splitting $E_F^n - E_F^p$ is
\begin{equation}\label{eq:S07}\begin{split}
    eV_\mathrm{oc} &= E_\mathrm{g} + E_\mathrm{U} \ln{\frac{np}{N_0^2}}\\
    &= E_\mathrm{g} + 2 E_\mathrm{U} \ln{\frac{n}{N_0}} , 
\end{split}\end{equation}
and the charge carrier density expressed in terms of $V_\mathrm{oc}$ and the effective band gap $E_\mathrm{g}$ is given by~\cite{kirchartz2011recombination,lange2013correlation,collins2016understanding,hofacker2017dispersive}
\begin{equation}\label{eq:S08}
    n = p = N_0 \cdot \exp{\left( -\frac{E_\mathrm{g} - eV_\mathrm{oc}}{2 E_\mathrm{U}}\right)} . 
\end{equation}

\subsubsection{Gaussian DOS} 
Supposing that the density of states of both electrons and holes is described by a gaussian, the charge carrier density is given by Eq.\,\eqref{eq:S05}. 
Assuming, for simplicity, that $\sigma_n = \sigma_p = \sigma$, and with $n = p$, the open-circuit voltage becomes 
\begin{equation}\label{eq:S09}\begin{split}
    eV_\mathrm{oc} &= E_\mathrm{g} - \frac{\sigma^2}{k_BT} + k_BT\ln{\frac{np}{N_0^2}} \\
    &= E_\mathrm{g} - \frac{\sigma^2}{k_BT} + 2 k_BT \ln{\frac{n}{N_0}} , 
\end{split}\end{equation}
which results in the charge carrier density given by~\cite{garcia2010temperature,lange2013correlation,hofacker2017dispersive}
\begin{equation}\label{eq:S10}
    n = p = N_0 \cdot \exp{\left(-\frac{E_\mathrm{g} - \sigma^2(k_BT)^{-1} - eV_\mathrm{oc}}{2k_BT}\right)} . 
\end{equation}

\subsubsection{Mixed DOS}

For the mixed DOS, where the DOS distributions of electrons and holes are different, we assume that electrons are described by the exponential and holes by the gaussian DOS. Their respective densities are defined by Eqs.\,\eqref{eq:S02} and \eqref{eq:S05}. 
As in the previous cases, with equal electron and hole densities $n = p$, the open-circuit voltage $V_\mathrm{oc}$ is given by 
\begin{equation}\label{eq:S11}\begin{split}
    eV_\mathrm{oc} &= E_\mathrm{g} - \frac{\sigma^2}{2k_BT} + E_\mathrm{U}\ln{\frac{n}{N_0}} + k_BT\ln{\frac{p}{N_0}}\\
    &= E_\mathrm{g} - \frac{\sigma^2}{2k_BT} + \left( E_\mathrm{U} + k_BT\right)\ln{\frac{n}{N_0}} , 
\end{split}\end{equation}
and the charge carrier density is~\cite{hofacker2017dispersive}
\begin{equation}\label{eq:S12}
    n = p = N_0 \cdot \exp{\left( -\frac{E_\mathrm{g} - \sigma^2(2k_BT)^{-1} - eV_\mathrm{oc}}{E_\mathrm{U} + k_BT}\right)} . 
\end{equation}
\vspace{0.3cm}

\subsection{The recombination rate}

In order to relate the recombination rate $R$ to the effective band gap $E_\mathrm{g}$ and the open-circuit voltage $V_\mathrm{oc}$, the expressions for $\theta$ and $n$ are used to replace the terms in Eq.\,(3). For the purely exponential DOS, we use Eqs.\,\eqref{eq:S03} and \eqref{eq:S08}, which yields~\cite{kirchartz2011recombination,hofacker2017dispersive,xiao2020relationship}
\begin{equation}\label{eq:S13}\begin{split}
    R &\approx k_r N_0^{1-\lambda} n^{\lambda + 1}\\
    &= k_r N_0^2 \cdot \exp{\left[ -\frac{E_\mathrm{g} - eV_\mathrm{oc}}{k_BT}\left(\frac{1}{2}+\frac{k_BT}{2E_\mathrm{U}}\right)\right]} . 
\end{split}\end{equation}

Supposing that the density of states of both electrons and holes is described by the gaussian DOS with the same $\sigma$, we substitute $\theta$ and $n$ in Eq.\,(3) by Eqs.\,\eqref{eq:S06} and \eqref{eq:S10}, respectively, and arrive at~\cite{burke2015beyond,hofacker2017dispersive}
\begin{equation}\label{eq:S14}\begin{split}
    R &\approx k_r n^2 \cdot \exp{\left(-\frac{\sigma^2}{2(k_BT)^2}\right)} \\
    &= k_r N_0^2 \cdot \exp{\left(-\frac{E_\mathrm{g} - \sigma^2(2k_BT)^{-1} - eV_\mathrm{oc}}{k_BT}\right)} . 
\end{split}\end{equation}

In our last scenario, electrons and holes are represented by different DOS functions. Without loss of generality, we describe recombination of electrons from the exponential and holes from the gaussian DOS. The total charge carrier concentration given by Eq.\,\eqref{eq:S12} replaces $n$ in Eq.\,(3), and the recombination rate expression is determined by $\theta$, hence the dominant recombination channel. If the channel involving recombination of mobile electrons from the exponential DOS with all holes from the gaussian DOS is dominant, then the trapping factor is defined by Eq.\,\eqref{eq:S03}, and the recombination rate is expressed as~\cite{hofacker2017dispersive}
\begin{equation}\label{eq:S15}\begin{split}
    R &\approx k_{r} N_0^{1-\lambda} n^{\lambda + 1}\\
    &= k_{r} N_0^2\cdot \exp{\left(-\frac{E_\mathrm{g} - \sigma^2(2k_BT)^{-1} - eV_\mathrm{oc}}{k_BT}\right)} , 
\end{split}\end{equation}
with the same result as for the purely gaussian DOS. If, on the other hand, recombination is controlled by mobile holes from the gaussian DOS recombining with all electrons from the exponential DOS, their trapping factor is defined by Eq.\,\eqref{eq:S06}, which yields~\cite{hofacker2017dispersive}
\begin{equation}\label{eq:S16}\begin{split}
    R &\approx k_{r} n^2 \cdot \exp{\left(-\frac{\sigma^2}{2(k_BT)^2}\right)}\\
    &= k_{r} N_0^2\cdot \exp{\left(-\frac{E_\mathrm{g} + \sigma^2(E_\mathrm{U} - k_BT)(2k_BT)^{-2} - eV_\mathrm{oc}}{\frac{1}{2}(E_\mathrm{U} + k_BT)}\right)} . 
\end{split}\end{equation}
Comparison of Eq.\,(1) to the expressions for the recombination rate given by Eqs.\,\eqref{eq:S13} to \eqref{eq:S16} yields Eqs.\,(4) and (5) in the main text.

\clearpage
\section{Light intensity-dependent ideality factors}

\begin{figure}[h]
    \centering
    \includegraphics[width=0.45\textwidth]{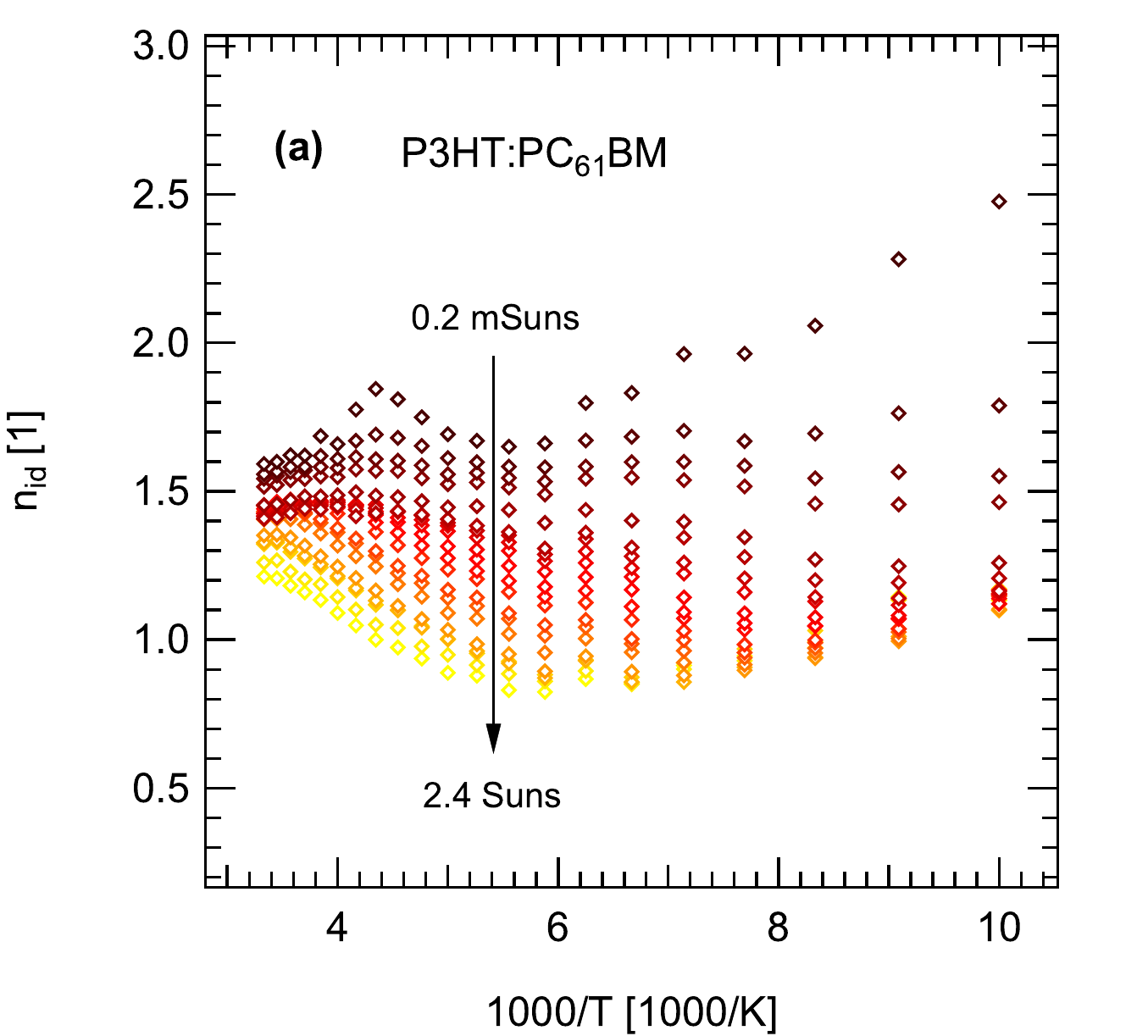}
    \includegraphics[width=0.45\textwidth]{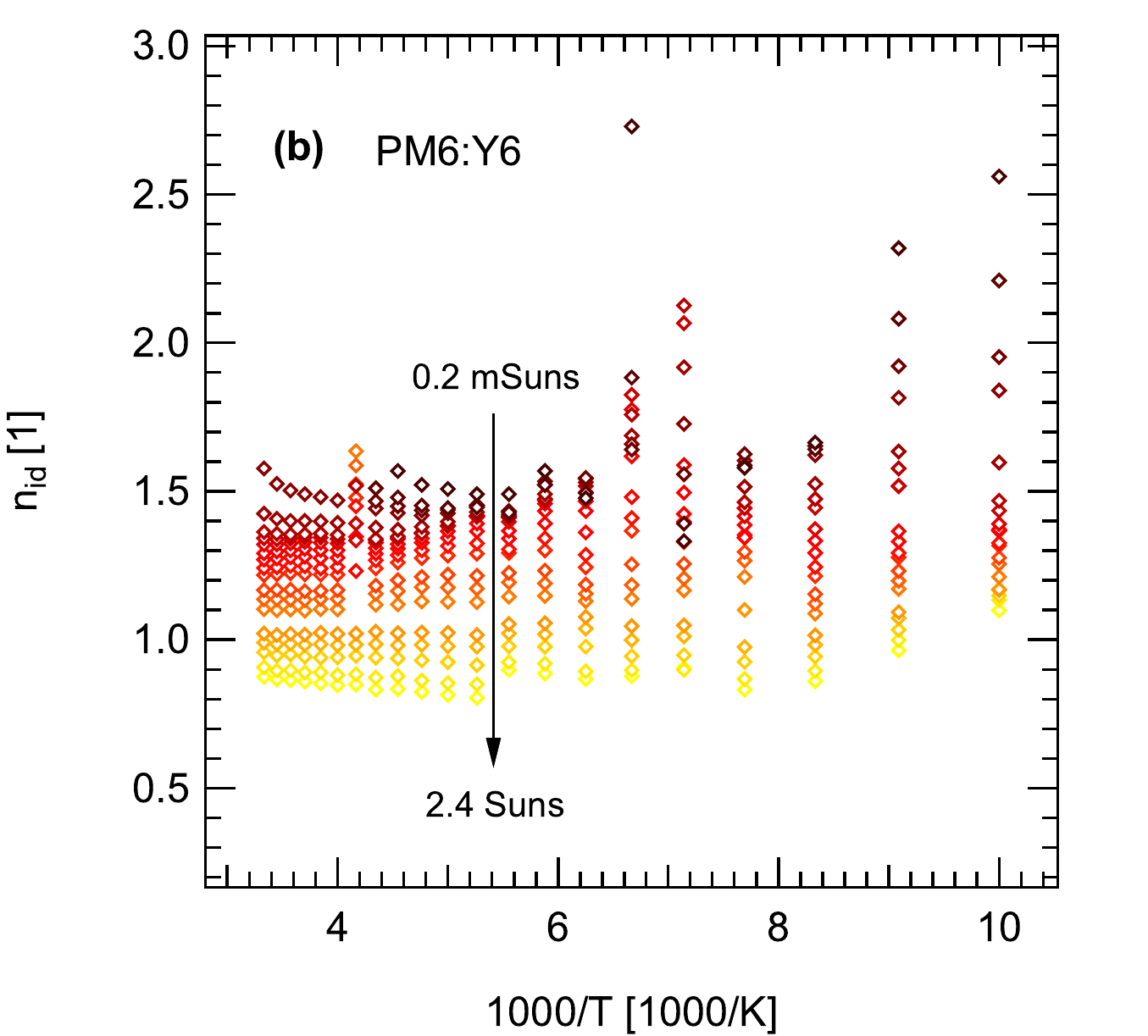}
    \includegraphics[width=0.45\textwidth]{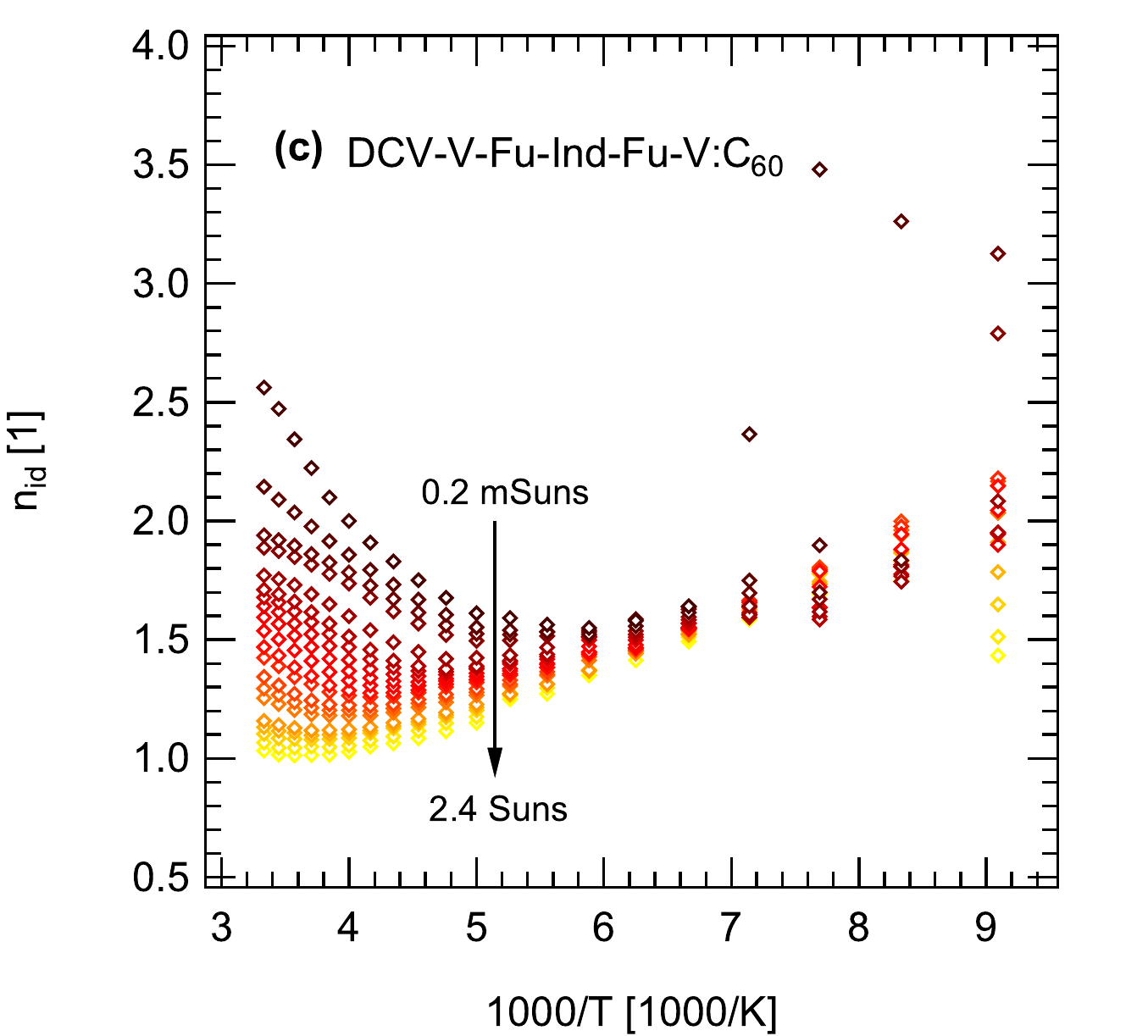}
    \caption{Light intensity-dependent ideality factors of (a) P3HT:PC$_{61}$BM, (b) PM6:Y6 and (c) DCV-V-Fu-Ind-Fu-V:C$_{60}$ as a function of the inverse temperature.}
    \label{fig:S03}
\end{figure}

The ideality factor of P3HT:PC$_{61}$BM at a given light intensity in Figure\,\ref{fig:S03}(a) is inversely correlated to $1/T$ between 170\,K and 300\,K. The recombination models described in the main text, on the other hand, predict whether constant $n_\mathrm{id} = 1$ or an increasing $n_\mathrm{id}$ with $1/T$. This apparent temperature dependence of $n_\mathrm{id}$ at a given light intensity deviates from the models due to the increasing quasi-Fermi splitting at lower temperatures. 
The same mismatch between the models and the apparent temperature dependence of $n_\mathrm{id}$ is observed for DCV-V-Fu-Ind-Fu-V:C$_{60}$ between 200\,K and 300\,K in Figure\,\ref{fig:S03}(c). Only at lower temperatures do the ideality factors of this system converge for different light intensities and follow the mixed DOS recombination model. 
The roughly temperature-independent $n_\mathrm{id}$ of PM6:Y6 in Figure\,\ref{fig:S03}(b) increases substantially at lower light intensities, and therefore can not be assigned to the recombination models characterised by $n_\mathrm{id} = 1$. All in all, analysing the temperature dependence of ideality factors at a given light intensity is not an effective strategy to draw conclusions about the dominant recombination mechanism present in a solar cell.
    
\bibliographystyle{apsrev4-2}
\bibliography{SI}